\documentstyle[12pt,amssymb,preprint,prd,aps,eqsecnum,epsf]{revtex}
\tighten

\begin{document}

\preprint{DAMTP-1998-105}
\title{Extending the Black Hole Uniqueness Theorems \\ I. Accelerating Black Holes: The Ernst Solution and $C$-Metric}
\author{Clive G. Wells\thanks{Electronic address: C.G.Wells@damtp.cam.ac.uk}}
\address{Department of Applied Mathematics and Theoretical Physics,
University of Cambridge, \\ Silver St., Cambridge CB3 9EW, United
Kingdom} 
\date{\today}
\maketitle

\begin{abstract}
We present black hole uniqueness theorems for the $C$-metric and Ernst solution. The proof follows a similar strategy as that used to prove the uniqueness of the Kerr-Newman solution, however the presence of an acceleration horizon provides some critical differences. We also show how to derive the Bunting/Mazur result (on the positivity of a suitable divergence required in the proof) using new methods. We briefly explain the importance of the uniqueness of the Ernst solution in relation to the proposed black hole monopole pair creation mediated by the related instanton.

\end{abstract}


\narrowtext

\def\nonum{\nonumber\samepage}
\relpenalty=10000\binoppenalty=10000

\def\be{\begin{equation}}
\def\ee{\end{equation}}
\def\beq{\begin{eqnarray}}
\def\eq{\end{eqnarray}}
 
\def\bar{\overline}
\def\bd{\b\delta}
\def\t{\otimes}
\def\h{\frc12}
\def\sh{\mbox{$\frac12$}}
\def\sq{\mbox{$\frac14$}}
\def\D{\nabla}
\def\frc#1#2{\displaystyle\frac{#1}{#2}}
\def\ip<#1,#2>{\langle#1,#2\rangle}

\def\v{{\vphantom{\dagger}}}
\def\Im{\mathop{\rm Im}\nolimits}
\def\sn{\mathop{\rm sn}\nolimits}
\def\cn{\mathop{\rm cn}\nolimits}
\def\dn{\mathop{\rm dn}\nolimits}
\def\sc{\sn\chi}
\def\cc{\cn\chi}
\def\dc{\dn\chi}
\def\cz{\chi_0}
\def\dd{\mbox{\boldmath $d$}}
\def\e#1{\mbox{\boldmath $e$}^#1}
\def\hd{\mbox{\boldmath$*$}}
\def\b#1{\mbox{\boldmath$#1$}}
\def\w{\wedge}
\def\k{\b k}
\def\m{\b m}
\def\lk{{\cal L}_K}
\def\lm{{\cal L}_m}
\def\({\left(}
\def\){\right)}

\def\ddd#1#2{\dd#1\otimes\dd#2}
\def\pp#1#2{\frc{\partial#1}{\partial#2}}
\def\ppp#1{\frac\partial{\partial#1}}
\def\Tr{{\rm Tr\,}}
\def\df#1{\hat#1}
\def\ska{\frc{\sqrt2}{\kappa^2A^2}}
\def\mii{{\cal M}_{\rm II}}
\def\gii{{\b g}_{{}_{\rm II}}}
\def\gtii{{\b {\tilde g}}_{{}_{\rm II}}}
 
\def\openone{\leavevmode\hbox{\small1\kern-3.8pt\normalsize1}}

\def\ignore{\gobble}
\def\endignore{\endgroup}
\long\def\gobble#1\endignore{}
 
\section{Introduction}

Over the last few years there has been considerable interest in the $C$-metric and related solutions as gravitational instantons. It is natural to try to ascertain the uniqueness (or otherwise) of these solutions as the instanton is only relevant if it dominates the euclidean path integral. Clearly the possibility of other instanton solutions could force us to re-examine the relevance of the physical process the original instanton is supposed to be mediating, in the current case that of black hole pair creation.

In this paper we give a complete proof of the uniqueness of the $C$-metric and Ernst solution. Much of the proof is carried over directly from the standard Kerr-Newman black hole uniqueness theorem (for a treatment in a style similar to  this paper see \cite{Heusler}). However at various points it is convenient to show the relationship with Riemann surfaces and complex manifolds. In particular the solutions we will be discussing have a nice relationship to Riemann surfaces that are topologically tori. This is related to the use of elliptic functions and integrals that will be a vital ingredient in our proof.

In Sect.~\ref{sect:em} we derive the Ernst parameterization and effective Lagrangian that is obtained upon dimensional reduction to three dimensions. The effective Lagrangian obtained a harmonic mapping Lagrangian whose metric is the Bergmann metric, we go on to show how to derive this metric from a projection from three-dimensional complex Minkowski space onto the unit ball. Once we have this construction, it is fairly straightforward to derive the Mazur/Bunting results, generalizing Robinson's identity, using these tools.

Given the construction of the Bergmann metric it is a simple matter to write down a number of internal symmetry transformations, in particular we look at the Harrison transformation which we apply to the $C$-metric in Sect.~\ref{sect:Cmetric}, this yields the Ernst solution. The $C$-metric and Ernst solution are then written in terms of elliptic functions, anticipating the introduction of such coordinates for a general spacetime representing an accelerating black hole.

In Sect.~\ref{sect:unique} we present the hypotheses for our uniqueness argument, and give a proof of the Generalized Papapetrou Theorem, we then justify the introduction of Weyl coordinates by making use of the Riemann Mapping Theorem. After explaining how the conformal factor of the two-dimensional metric can be determined, we present boundary conditions for both the $C$-metric and Ernst solution that complete the proof of the uniqueness theorems.

\section{The Effective Three Dimensional Theory}
\label{sect:em}
In this section we investigate the effective Lagrangian arising from
Einstein-Maxwell theory after a dimensional reduction on a timelike Killing vector field
 $K=\partial/\partial t$. Our starting point is to write the metric in the form
\be
\b g=-V\(\dd t+\b C\)\otimes\(\dd t+\b C\)+V^{-1}\gamma_{ij}\ddd{x^i}{x^j}
\label{eq:mtrc}
\ee
and introduce the twist 1-form $\b\omega$ associated with the vector field $K$. We set
\be
\b\omega=\hd(\k\w\dd\k)\ ,
\ee
where $\k=-V\(\dd t+\b C\)$ is the 1-form associated with $K$, and $\hd$ is the Hodge dual associated with the four dimensional metric. Acting on a $p$-form twice we have $\hd\(\hd\b\omega^{(p)}\)=(-1)^{p+1}\b\omega^{(p)}$. Let us define $\b H=\dd\b C$.  We have therefore
\be
\b H=\dd\(\frc\k{i_K\k}\)
\label{eq:defH}
\ee
where $i_K$ is the antiderivation on forms, defined by taking the inner product with $K$. It can also be defined in terms of the Hodge dual by $i_K\b\omega^{(p)}=-\hd\(\k\w\hd\b\omega^{(p)}\)$ for any $p$-form, $\b\omega^{(p)}$. Using these  facts we find
\be
\frc{\hd\b\omega}{i_K\k}=\k\w\b H\ .
\ee
The Lie derivative is defined to be $\lk=\dd i_K+i_K\dd$. Killing's equation implies $\lk\k=0$. Using this and Eq.~(\ref{eq:defH}), we note that $i_K\b H=0$. Thus,
\beq
\frc{\hd\(\b\omega\w\hd\b\omega\)}{\(i_K\k\)^2}&=&
\hd\(\hd\(\k\w\b H\)\w\k\w\b H\)\nonum\\
&=&\hd\(i_K\hd\b H\w\k\w\b H\)\nonum\\
&=&-\(i_K\k\)\hd\(\b H\w\hd\b H\).
\eq
In index notation, and in terms of the three metric $\gamma_{ij}$ this result states:
\be
\frc{2\omega_i\omega_j\gamma^{ij}}{V^4}=H_{ij}H_{kl}\gamma^{ik}\gamma^{jl}.
\ee

We now introduce the electromagnetic field. The Lagrangian density may be written
as
\be
{\cal L}=\sqrt {|g|}\(R-F_{ab}F^{ab}\)\ .
\label{eq:lagfdensity}
\ee
The field $\b F$ being derived from a vector potential: $\b F=\dd\b A$.
 We will be assuming that the Maxwell field obeys the appropriate symmetry
 condition: $\lk\b F=0$. The exactness of $\b F$ implies that
\be
\dd i_K\b F=0.
\ee
It is now convenient to introduce the electric and magnetic fields by
\be
\b E=-i_K\b F\qquad\mbox{and}\qquad\b B=i_K\hd\b F\ .
\ee
Notice that $i_K\b E=i_K\b B=0$ as a consequence of the general result $(i_K)^2=0$.
It will be convenient to decompose the electromagnetic field tensor in terms of the newly defined
electric and magnetic fields as follows:
\be
\b F=\frc{-\k\w\b E-\hd\(\k\w\b B\)}{i_K\k}\ .
\ee
The Lagrangian for the electromagnetic interaction, proportional to
 $F_{ab}F^{ab}$, may be written in terms of the $\b E$ and $\b B$ fields,
\beq
\sh F_{ab}F^{ab}&=&-\hd\(\b F\w\hd\b F\)\nonum\\
&=&\frc{-\hd\(\b E\w\hd\b E-\b B\w\hd\b B\)}{i_K\k}\nonum\\
&=&\frc{|\bf B|^2-|\bf E|^2}{V}\ .
\eq
Two of Maxwell's equations, namely the ones involving the divergence 
of $\bf B$ and the curl of $\bf E$ arise not from the Lagrangian (\ref{eq:lagfdensity}), but 
rather from the exactness of $\b F$. In order to find these equations 
we evaluate $i_K\hd\dd\b F=0$:
\beq
i_K\hd\dd\b F&=&-i_K\left[\hd\(\dd\(\frc\k{i_K\k}\)\w\b E\)
+\bd\(\k\w\(\frc{\b B}{i_K\k}\)\)\right]
\nonum\\
&=&(i_K\k)\,\bd\(\frc{\b B}{i_K\k}\)+\frc{i_E\b\omega}{i_K\k}
\eq
or in vector notation
\be
\D.\(\frc{\bf B}V\)+\frc{\b{\b\omega}.{\bf E}}{V^2}={\bf0}\ .
\ee
 This equation is a constraint on the fields. We also notice that
 $\dd\b F=0$ implies together with the symmetry condition, that
 $i_K\dd\b F=-\dd i_K\b F=-\dd\b E=0$ and hence that locally we may write 
 $\b E=-\dd\Phi$. In order to progress we will also need to 
know about the divergence of $\b\omega/V^2$. Firstly observe that
\be
\dd\(\frc\k{i_K\k}\)=i_K\b D
\ee
for some 3-form $\b D$. This follows from the fact that when
 we apply $i_K$ to  the 2-form on the left hand side we get zero
 (as previously mentioned). Decomposing the 
left hand side into `electric' and `magnetic' parts we see that 
the `electric' part is zero. This leads to
\beq
-\D.\(\frc{\b\omega}{V^2}\)=\bd\(\frc{\b\omega}{V^2}\)
&=&\hd\left[\dd\(\frc\k{i_K\k}\)\w\dd\(\frc\k{i_K\k}\)\right]\nonum\\
&=&\hd(i_K\b D\w i_K\b D)\nonum\\
&=&\hd i_K(\b D\w i_K\b D)=0,
\eq
as the last equation involves the inner product of a 5-form, which 
automatically vanishes. The constraint equation may therefore 
be written as a divergence:
\be
\D.\(\frc {\bf B}V-\frc{\b\omega\Phi}{V^2}\)={\bf0}\ .
\ee
In order to impose the constraint we make use of a Legendre transformation.
 To this end we introduce a Lagrangian multiplier, $\Psi$. 
After discarding total divergences the Lagrangian to vary is given by 
\be
{\cal L}=\sqrt{|g|}\(R+2\left[\frc{|\D\Phi|^2-{\bf B}^2}V+\frc{2\,{\bf B}.\D\Psi}V
-\frc{\b\omega.(\Phi\D\Psi-\Psi\D\Phi)}{V^2}\right]
\).
\ee
Varying with respect to $\bf B$ we conclude that ${\bf B}=\D\Psi$ and
performing the dimensional reduction to three dimensions we find
\be
{\cal L}=\sqrt{\gamma}\({}^3\!R-2\left[\frc{|\D V|^2-|\b\omega|^2}{4\,V^2}-\frc{|\D\Phi|^2+|\D\Psi|^2}V
+\frc{\b\omega.(\Phi\D\Psi-\Psi\D\Phi)}{V^2}\right]
\)\ .
\label{eq:lem1}
\ee
All indices in the above equation are raised and lowered using $\gamma_{ij}$ and its inverse. We mention that $\b\omega$ may not be varied freely in the above Lagrangian. It is constrained by the requirement $\bd\(\b\omega/V^2\)=0$. In order to impose this constraint we introduce another Lagrangian multiplier, $\omega$. After discarding a total divergence the new lagrangian reads,
\be
{\cal L}=\sqrt{\gamma}\({}^3\!R-2\left[\frc{|\D V|^2-|\b\omega|^2}{4\,V^2}-\frc{|\D\Phi|^2+|\D\Psi|^2}V
+\frc{\b\omega.(\D\omega+2(\Phi\D\Psi-\Psi\D\Phi))}{2V^2}\right]
\)\ .
\label{eq:lem}
\ee
Varying with respect to $\b\omega$ yields,
\be
\b\omega=\D\omega+2(\Phi\D\Psi-\Psi\D\Phi).
\ee
Now substitute this back, we find
\be
{\cal L}=\sqrt{\gamma}\({}^3\!R-2\left[\frc{|\D V|^2+|\b\omega|^2}{4\,V^2}-\frc{|\D\Phi|^2+|\D\Psi|^2}V
\)\right] 
\label{eq:lem2}
\ee
where the 3-vector $\b\omega_i$ is defined to be $\D\omega+2(\Phi\D\Psi-\Psi\D\Phi)$ in this formula.

 It turns out to be highly useful to combine the two potentials, 
$\Phi$ and $\Psi$ into a single complex potential, $\psi=\Phi+i\Psi$. 
Using this complex potential we have
\be
i\b\omega=i\dd\omega-\psi\dd\bar\psi+\bar\psi\dd\psi\ .
\ee

We are now in a position to define the Ernst potential
 \cite{Ernstpot}, $\epsilon$ by, $\epsilon=V-|\psi|^2+i\omega$. Then clearly
\be
\dd\epsilon+2\bar\psi\dd\psi=\dd V+i\b\omega\ .
\ee
Substituting this into Eq.~(\ref{eq:lem2}) we get
\beq
{\cal L}&=&\sqrt{|\gamma|}\({}^3\!R
-2\left[\frc{|\D\epsilon+2\bar\psi\D\psi|^2}{4V^2}
-\frc{|\D\psi|^2}V\right]\)\\
&=&\sqrt{|\gamma|}\({}^3\!R-2\gamma^{ij}G_{AB}\pp{\phi^A}{x^i}\pp{\phi^B}{x^j}\)
\qquad\qquad\(\phi^A\)=\pmatrix{\epsilon\cr\psi\cr}\ .
\eq
with the harmonic mapping target space metric $G_{AB}$ given by
\be
\b G=G_{AB}\ddd{\phi^A}{\phi^B}=
\frc{\(\dd\epsilon+2\bar\psi\dd\psi\)\otimes_{{}_S} \(\dd{\bar\epsilon}+2\psi\dd{\bar\psi}\)}
{4V^2}-\frc{\dd\psi\otimes_{{}_S}\dd{\bar\psi}}{V}\ .
\ee
Here, $\otimes_{{}_S}$ is the symmetrized tensor product. We now state the analogous result for the situation when using a spacelike Killing vector, $m=\partial/\partial\phi$, In fact it merely replaces $V$ by $-X$ throughout, where $X=i_m\m$. This can be seen by considering $\b\gamma\mapsto -\b\gamma$, $V\mapsto-X$ and $t\mapsto\phi$ in Eq.~(\ref{eq:mtrc}). It turns out that for the black hole uniqueness result, the formulation relying on the angular Killing vector is more useful. Explicitly then,
\be
\b G=G_{AB}\ddd{\phi^A}{\phi^B}=
\frc{\(\dd\epsilon+2\bar\psi\dd\psi\)\otimes_{{}_S} \(\dd{\bar\epsilon}+2\psi\dd{\bar\psi}\)}
{4X^2}+\frc{\dd\psi\otimes_{{}_S}\dd{\bar\psi}}{X}\ .
\ee
 These metrics are conveniently written in terms of new variables with
\be
\xi=\frc{1\pm\epsilon}{1\mp\epsilon};\qquad\qquad\qquad\eta=
\frc{2\psi}{1\mp\epsilon}.
\label{eq:eppsi}
\ee
Here and in the following equations the top sign corresponds to the spacelike Killing vector reduction whilst the lower one to a reduction performed on a timelike Killing vector. The metric $\b G$ then takes the form
\be
\b G=\frc{\(1\mp|\eta|^2\)\dd\xi\otimes_{{}_S}\dd{\bar\xi}+\(1-|\xi|^2\)\dd\eta\otimes_{{}_S}\dd{\bar\eta}
\pm\xi\bar\eta\dd{\bar\xi\otimes_{{}_S}}\dd\eta
\pm\bar\xi\eta\dd\xi\otimes_{{}_S}\dd{\bar\eta}}{\(1-|\xi|^2\mp|\eta|^2\)^2}\ .
\label{eq:berg}
\ee
The metric with the upper signs is the Bergmann metric, and it is the natural generalization of the
 Poincar\'e metric to higher dimensions, as we shall see in the following 
sections there is an important $SU(1,2)$ action preserving this metric. 
With these transformations we will be able to generate new solutions from old. It will also allow us to prove a particular identity vital for establishing the uniqueness theorems we are interested in.

\subsection{The Poincar\'e and Bergmann Metrics}
\label{sect:Poincare}

The Poincar\'e and Bergmann metrics have simple geometrical constructions. 
The Poincar\'e metric is the natural metric to put on the unit disc, as its isometries are
 precisely those M\"obius maps that leave the unit disc 
invariant. 

To construct these metrics our starting point is the vector space ${\Bbb C}^{n+1}$.
 For the Poincar\'e metric, $n=1$, whilst for the Bergmann metric $n=2$. 
 We will be using 
complex coordinates $z_0,z_1,...,z_n$. Let us write
\be
\eta=\pmatrix{-1&&&&\cr
                &1&&&&\cr
                &&1&&&\cr
                &&&\ddots&\cr
                &&&&1\cr}\qquad\mbox{or}\qquad
\eta=\pmatrix{1&&&&\cr
                &-1&&&&\cr
                &&1&&&\cr
                &&&\ddots&\cr
                &&&&1\cr}
\ee
The first case will be relevant for the spacelike Killing vector reduction, whilst the second is that required when discussing the timelike case.

We define an indefinite inner product using $\ip<w,z>=w^\dagger\eta z$. We therefore 
have $\dd\b s=\|\dd z\|$. This metric induces metrics on the hyperboloids defined by
\be
\|z\|^2=\mp1.
\label{eq:hypeq}
\ee 
 We will find it convenient to write
\be
\pmatrix{z_0\cr z_1\cr\vdots\cr z_n\cr}=re^{it}\pmatrix{1\cr v_1\cr\vdots \cr
v_n\cr},\quad\quad\quad r^{-2}=1\mp\|v\|^2.
\ee
In addition we put the natural induced metric on the space of vectors $v$, given by its embedding as the hyperplane $z_0=0$ in ${\Bbb C}^{n+1}$ where $z_i=v_i$ for all $i=1,\ldots,n$. 

We quickly
establish that
\be
{\dd\b s}^2=\dd z^\dagger\eta\dd z=\mp\left[\dd t+\b A\right]^2+r^2\|\dd v\|^2\pm r^4|\ip<v,\dd v>|^2,
\ee
and
\be
{\b A}=\mp{i\over2}r^2\(\ip<v,\dd v>-\ip<\dd v,v>\).
\ee
We also notice that $\dd t+{\b A}$ may be written as
\be
\dd t+\b A=\mp{i\over2}\(\dd z^\dagger\eta z-z^\dagger\eta\dd z\).
\ee
Hence we find that the metric can then be expressed as
\be
{\dd\b s}^2=\mp\left[\dd t+\b A\right]^2+g_{a\dot b}\dd v^a\dd \bar
v^{\dot b}.
\ee
We call $g_{a\dot b}$ the Bergmann metric for $n=2$ and with the upper signs, (or the Poincar\'e metric
for $n=1$). We remark in passing that these metrics are K\"ahler, and
as with all K\"ahler metrics we may derive the metric and symplectic 
2-form from a K\"ahler potential,
$K$, where in this case $K=\pm\log r$. 
 
We will now proceed to investigate the isometries of the hyperboloids with these metrics, by
acting with elements of $U(1,n)$. We notice that if constant $g\in U(1,n)$ (so that
$g^\dagger\eta g=\eta$) then we will have that
\be
\dd t+\b A\mapsto\mp{i\over2}\(\dd z^\dagger g^\dagger\eta gz-z^\dagger
g^\dagger\eta g\dd z\)=\mp{i\over2}\(\dd z^\dagger\eta z-z^\dagger\eta\dd z\)
=\dd t+\b A.
\ee
So the elements of $U(1,n)$ will generate isometries of the Bergmann metric. When we project onto the domain $\|v\|^2<1$ (spacelike) or $\|v\|^2>-1$ (timelike) we get
isometries from the elements of $SU(1,n)$, as a mere change of phase gives rise
to the same isometry of the Bergmann metric, it just generates a translation of
the $t$-coordinate.
The group of isometries acts transitively on the domain, and hence we
draw the conclusion that the curvature must be constant. It is useful to look at the
stabilizer of some point, for simplicity (and without loss of generality) 
we take the origin. We observe that
if
\be
g\pmatrix{1\cr0\cr\vdots\cr0}=
re^{it}\pmatrix{1\cr0\cr\vdots\cr0},\qquad\quad g\in SU(1,n),
\ee
then $g\in S(U(1)\times U(n))$ in the spacelike case, so we may identify the domain with the (symmetric) space
of left cosets $SU(1,n)/S\(U(1)\times U(n)\)$. Similarly, for the timelike case we find the symmetric space $SU(1,n)/S\(U(1)\times U(1,n-1)\)$.

In this section we have seen how to construct the Bergmann metric in terms of a suitable
projection and an auxiliary complex vector space. The Einstein-Maxwell system can
be expressed in this language by defining
\be
z=\pmatrix{1\mp\epsilon\cr1\pm\epsilon\cr2\psi}
\label{eq:electropara}
\ee
with the Lagrangian written as
\be
{\cal L}=\sqrt\gamma\({}^3\!R\pm2\frc{\|\(\D z\)^\perp\|^2}{\|z\|^2}\).
\label{eq:ellag}
\ee
We have defined the orthogonal component of a vector quantity $A$ as $A^\perp=A-\ip<z,A>z/\|z\|^2$.
It is clear that from its construction there is much symmetry in this system and we will be exploiting this fact in the
next few sections.

\subsection {The Divergence Identity.}
\label{sect:divergence}

In this section we give a new proof of the positivity of the electromagnetic
generalization of Robinson's identity.  In contrast to the proofs given by Bunting \cite{Bunting} and Mazur \cite{Mazur,Mazur2} we do not
lean too heavily on the sigma-model formalism but rather use the complex
variable embedding of the hyperboloid in complex Minkowski space given
in the previous section. For this section we will work exclusively in the case of a reduction on a spacelike Killing vector.

It should be noticed that there is some gauge freedom in the above 
Lagrangian (\ref{eq:ellag}); specifically it is unchanged if we multiply 
$z$ by an arbitrary complex
 function. This is a reflection of the fact the construction of the Bergmann
 metric as a projection of complex lines. Our Lagrangian gives rise to the following 
 equation of motion:
\be
\D(\D z)^\perp=\(\frc{-1}{\|z\|^2}\)\(\|(\D z)^\perp\|^2+\ip<z,\D z>(\D z)^\perp\).
\ee
When we come to apply this result we will be using the Ernst potentials derived from the angular Killing vector $\partial/\partial\phi$. In consequence the three metric $\b\gamma$ will be indefinite and we shall show it may be written as
\be
\b\gamma=-\rho^2\ddd tt+\Sigma\(\ddd\rho\rho+\ddd zz\).
\ee
Our proof of the positivity of the divergence quantity to be introduced shortly involves terms such as $\ip<\gamma^{ab}\D_a z,\D_b z>$ etc. The indefiniteness of the metric will not be a problem, as nothing depends on $t$, and the metric only appears when contracting the gradient operator.

The equation of motion implies the expression:
\be
(\D^2z)^\perp=0. \qquad\mbox{i.e.,}\qquad\D^2z=\(\frc1{\|z\|^2}\)
\ip<z,\D^2z>z.
\label{eq:fieldeq}
\ee
As yet we have not made use of our gauge freedom. To begin with we 
shall use the
 freedom we have to normalize $z$ so that $\|z\|^2=-1$. In addition 
 we still have
 the freedom to multiply $z$ by an arbitrary phase. At any point we can 
exploit this freedom to set $\ip<z,\D z>=0$. However, its derivative
 will not vanish in general. The normalization we have imposed 
implies $\D^2\|z\|^2=0$. Consequently we have,
\be
\ip<z,\D^2z>+\ip<\D^2z,z>=-2\|\D z\|^2=-2\|(\D z)^\perp\|^2.
\ee
The last equality coming from the phase gauge condition.
 Henceforth we will always impose these two conditions and 
 therefore $\D z=(\D z)^\perp$ at the point $z$.

The divergence identity comes from examining the Laplacian of
\be
S=-\|z_1\w z_2\|^2,
\ee
where we have extended the inner product to the exterior
 algebra in the standard manner. The fields $z_1$ and $z_2$
 are assumed to obey both the field equation and the gauge conditions. We
 might notice that $S$ is invariant under arbitrary changes 
in phase of $z_1$ and $z_2$. For the moment we point out that the imposition
 of our phase gauge condition merely serves to make our 
calculations simpler: the expansion of $S$ does not depend on
 the parallel component of $\D z$.

Before we perform the calculation we make the useful observation,
\beq
\|z_1\w z_2\|^2&=&1-|\ip<z_1,z_2>|^2\nonum\\
               &=&-\|z_1^{\perp_2}\|^2\le0.
\eq
Where $z_1^{\perp_2}$ is the projection of $z_1$ orthogonal to $z_2$, as such it is orthogonal 
to a timelike vector, and is therefore spacelike.

Evaluating $\D^2S$ we find,
\beq
\D^2S&=&-\ip<\D^2z_1\w z_2+2\D z_1\w\D z_2+z_1\w\D^2 z_2,z_1\w z_2>
\nonum\\&&{}-\ip<z_1\w z_2,\D^2z_1\w z_2+2\D z_1\w\D z_2+z_1\w\D^2 z_2>
\nonum\\&&{}-2\|\D(z_1\w z_2)\|^2.
\eq
Making use of Eq. (\ref{eq:fieldeq}) we find
\beq
\D^2S&=&-2\|z_1\w z_2\|^2(\|\D z_1\|^2+\|\D z_2\|^2)-2\|\D(z_1\w z_2)\|^2
\nonum\\&&{}-2\ip<\D z_1\w\D z_2,z_1\w z_2>-2\ip<z_1\w z_2,\D z_1\w\D z_2>
\nonum\\&=&2|\ip<z_1,\D z_2>+\ip<\D z_1,z_2>|^2
\nonum\\&&+2|\ip<z_1,z_2>|^2(\|\D z_1\|^2+\|\D z_2\|^2)
\nonum\\&&+2\ip<z_1,z_2>\ip<\D z_2,\D z_1>+2\ip<z_2,z_1>\ip<\D z_1,\D z_2>.
\eq
Next we define $\Omega=\D(z_1\w z_2)$ and evaluate the norm of the following 
quantities,
\be
\ip<z_1,\Omega>=-(\D z_2+\ip<z_1,z_2>\D z_1+\ip<z_1,\D z_2>z_1)
\ee
and
\be
\ip<z_2,\Omega>=\D z_1+\ip<z_2,z_1>\D z_2+\ip<z_2,\D z_1>z_2.
\ee
Notice that by construction each is spacelike, being orthogonal to the timelike vectors $z_1$ and $z_2$ respectively.
We find that
\beq
\|\ip<z_1,\Omega>\|^2+\|\ip<z_2,\Omega>\|^2&=&
\(1+|\ip<z_1,z_2>|^2\)\(\|\D z_1\|^2+\|\D z_2\|^2\)\nonum\\
&&{}+|\ip<z_1,\D z_2>|^2+|\ip<z_2,\D z_1>|^2\nonum\\&&{}
+2\ip<z_1,z_2>\ip<\D z_2,\D z_1>+2\ip<z_2,z_1>\ip<\D z_1,\D z_2>.\hspace{2cm}
\eq
Hence
\beq
\D^2S&=&\|\ip<z_1,\Omega>\|^2+\|\ip<z_2,\Omega>\|^2+|\D\ip<z_1,z_2>|^2\nonum\\&&{}
+\(|\ip<z_1,z_2>|^2-1\)\( \|\D z_1\|^2+\|\D z_2\|^2\)\nonum\\&&{}
+\ip<\D z_1,z_2>\ip<z_1,\D z_2>+\ip<\D z_2,z_1>\ip<z_2,\D z_1>.
\eq
It only remains to notice that
\beq
\left|\ip<\D z_1,z_2>\ip<z_1,\D z_2>+\ip<\D z_2,z_1>\ip<z_2,\D z_1>\right|&\le& 2|\ip<\D z_1,z_2>|\,|\ip<z_1,\D z_2>|\nonum\\
&\le&2\|z_1^{\perp_2}\|\,\|z_2^{\perp_1}\|\,\|\D z_1\|\,\|\D z_2\|\nonum\\
&\le&\(|\ip<z_1,z_2>|^2-1\)\(\|\D z_1\|^2+\|\D z_2\|^2\)\ .\nonum\\ 
\eq
We have made use of the Cauchy-Schwarz inequality on the 
positive-definite subspaces
 orthogonal to $z_1$ and to $z_2$ together with the AM-GM inequality. 
Putting all this together we have therefore shown that
\be
\D^2S\ge0.
\ee
We have equality if and only if $\|z_1\w z_2\|$ is constant. 
In particular if $z_1$ and $z_2$ agree up to a phase anywhere then the 
constant is zero.

Returning to the Ernst parameterization, we set
\be
z=\frc1{2\sqrt X}\pmatrix{1-\epsilon\cr1+\epsilon\cr2\psi}
\ee
with the Ernst potentials derived from an angular Killing vector. It is the angular Killing vector rather than the timelike one that plays an important r\^ole in the black hole uniqueness theorems. We have
\beq
\epsilon&=&-X-|\psi|^2+iY,\\
\psi&=&E+iB.
\eq
It is worth pointing out that labelling the potentials $E$ and $B$ is conventional, however they are electric and magnetic potentials with respect to the angular Killing vector. In consequence $E$ actually determines the magnetic field as it is physically understood, and $B$ determines the electric field.

The condition $\|z_1\w z_2\|=0$ becomes
\be
\frc{{\df X}^2+2(X_1+X_2)\({\df E}^2+{\df B}^2\)+\({\df E}^2+{\df B}^2\)^2+
\(\df Y+2E_1B_2-2B_1E_2\)^2}{4X_1X_2}=0.
\ee
where we have used the abbreviation ${\df A}=A_2-A_1$. 
Accordingly $X_1=X_2$, $Y_1=Y_2$, $E_1=E_2$ and $B_1=B_2$. The equality of these quantities will be sufficient to establish the uniqueness of the solution as a whole.

\subsection{The Internal Symmetry Transformations for Einstein-Maxwell Theory}

We have seen how the Lie group $SU(1,2)$ acts on the complex hyperboloid $\|z\|={\rm constant}$, we now detail these transformations explicitly, these transformations are the Kinnersley group \cite{Kinnersley}.
It is highly useful to define the involutive automorphism $\sigma:SU(1,2)\rightarrow SU(1,2)$ by $\sigma(A)=\eta A\eta$. We are already aware of some of the isometries of the Bergmann metric, for instance we may add a constant to the twist potential:
\beq
\epsilon&\mapsto&\epsilon+it,\nonum\\{}
\psi&\mapsto&\psi,
\eq
where $t$ is real. This transformation corresponds to the matrix
\be
A=\pmatrix{1-it/2&-it/2&0\cr
           it/2&1+it/2&0\cr
           0&0&1}\ .
\ee
with action defined by
\be
r'e^{it'}\pmatrix{1\cr\xi'\cr\eta'}=re^{it}A\pmatrix{1\cr\xi\cr\eta}.
\ee
The matrix $\sigma(A)$ generalizes the Ehlers' transformation \cite{Ehlers}:
\beq
\epsilon&\mapsto&\frc\epsilon{1+it\epsilon}\ ,\nonum\\
\psi&\mapsto&\frc\psi{1+it\epsilon}\ .
\eq
Another obvious isometry of the Bergmann metric results from making gauge 
transformations to the electric and magnetic potentials:
\beq
\epsilon&\mapsto&\epsilon-2\bar\beta\psi-|\beta|^2,\\
\psi&\mapsto&\psi+\beta.
\eq
This arises from considering the $SU(1,2)$-matrix
\be
B=\pmatrix{1+|\beta|^2/2&|\beta|^2/2&\bar\beta\cr-|\beta|^2/2&1-|\beta|^2/2&
-\bar\beta\cr\beta&\beta&1},\quad\quad\beta\in{\Bbb C}.
\ee
The matrix $\sigma(B)$ gives rise to the Harrison transformation \cite{Harrison}:
\beq
\epsilon&\mapsto&\Lambda^{-1}\epsilon,\nonum\\
\psi&\mapsto&\Lambda^{-1}(\psi+\beta\epsilon),\nonum\\
\Lambda&=&1-2\bar\beta\psi-|\beta|^2\epsilon.
\label{eq:ht1}
\eq
The Harrison transformation is what is required to move from the $C$-metric to the Ernst solution. The uniqueness of these two solutions being the primary concern in this paper.

Finally to complete a set of eight generators for the group 
consider the combined rescaling of the Killing vector and electromagnetic duality rotation:
\beq
\epsilon&\mapsto&|\alpha|^2\epsilon,\nonum\\ 
\psi&\mapsto&\alpha\psi\qquad\alpha\in{\Bbb C}.
\eq
which corresponds to the matrix
\be
C=\pmatrix{\(\alpha^{-1}+\bar\alpha\)/2&\(\alpha^{-1}-\bar\alpha\)/2&0\cr
           \(\alpha^{-1}-\bar\alpha\)/2&\(\alpha^{-1}+\bar\alpha\)/2&0\cr
           0&0&1}.
\ee
The matrix $\sigma(C)$ corresponds to a redefinition 
of the parameter $\alpha$ and hence does not give rise to any new transformations.

\section{The Charged $C$-metric}
\label{sect:Cmetric}

The Vacuum $C$-metric has a history going back as far as 1918 \cite{Levi}, its electromagnetic
generalization was discovered in 1970 by Kinnersley and Walker \cite{Kinner}. It might be
noted however that this generalization is not simply a Harrison Transformation
on the timelike Killing vector, as is the case for charging up the Schwarzschild
solution to get the Riessner-Nordstr\o{}m black hole. In Sect.~\ref{sect:melmag} we
will be applying the Harrison transform to the charged $C$-metric but using the
{\em angular\/} Killing vector -- This is Ernst's solution. To begin with we
review the solution
determined by Kinnersley and Walker:
\be
\b g=r^2\({{\dd x\otimes\dd x}\over G(x)}-{{{\bf
d}y\otimes\dd y}\over G(y)}+G(x)\dd \alpha\otimes\dd \alpha+G(y){\bf
d}t\otimes\dd t\)
\ee
with
\beq
Ar&=&(x-y)^{-1},\\
G(x)&=&1-x^2-2\tilde mx^3-\tilde g^2x^4,\\
\tilde m&=&mA\qquad\mbox{and}\qquad\tilde g=gA.
\eq

The case $g=0$ is the vacuum solution. If we
take the limit $A\rightarrow0$, we discover that the solution reduces to the
Riessner-Nordstr\o{}m solution where $m$ and $g$ are the mass and charge of the
black hole. We remark that $m$ is not the ADM mass (unless $A=0$). The
ADM mass is zero, (the ADM 4-momentum is invariant under boosts and
rotations and therefore must be zero). The quantity $A$ is the acceleration of the
world-line $r=0$ when $m$ and $g$ are zero. We conclude that the $C$-metric
represents an accelerating black hole. The charged $C$-metric has a conical
singularity running along the axis. We can arrange by choosing the periodicity of the angular coordinate appropriately to eliminate this conical defect from one part of the axis. 

\begin{figure}[ht]
{\leavevmode\epsfxsize\hsize\epsfysize8cm\epsfbox{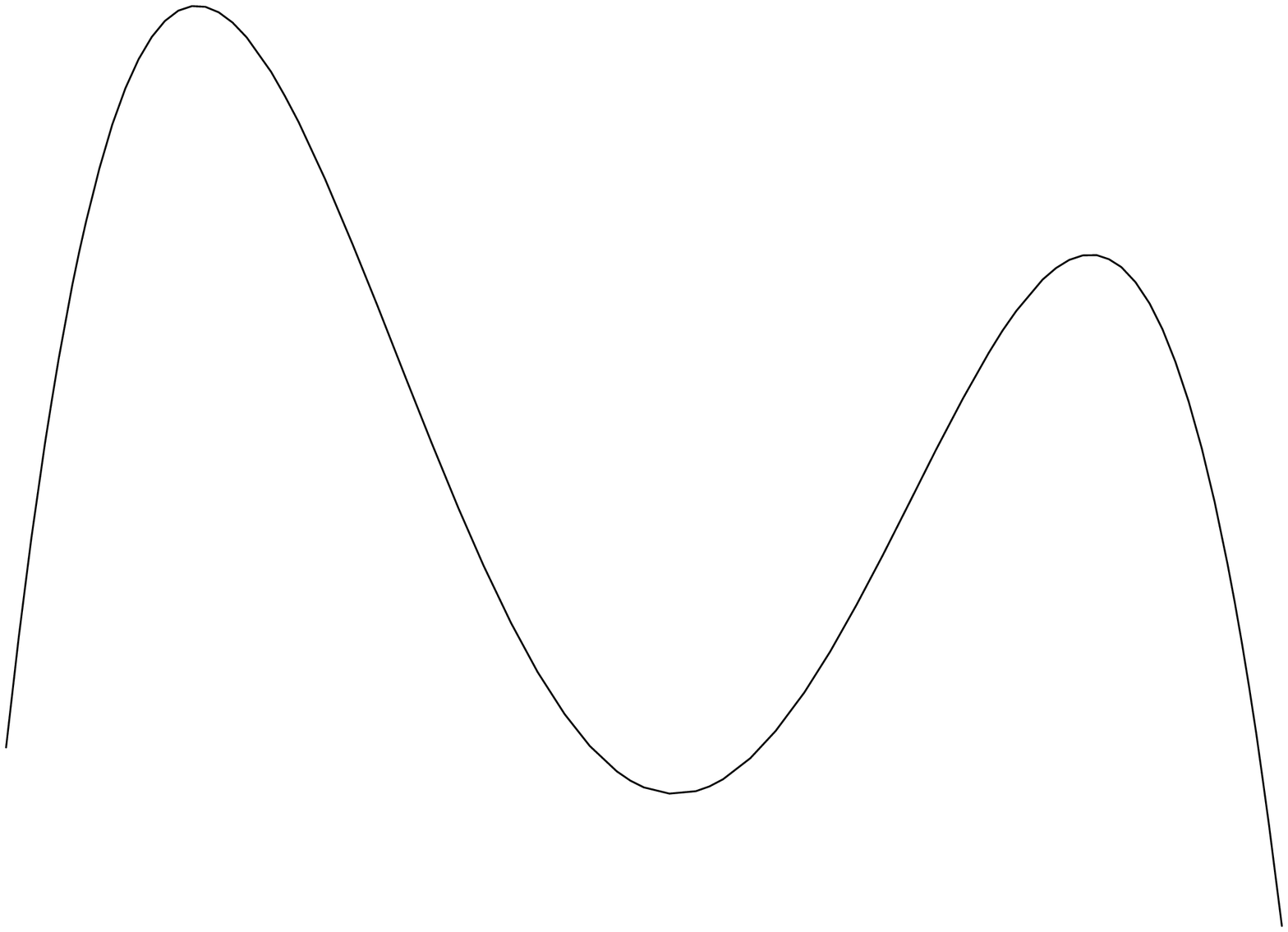}}
\begin{picture}(100,-600)
\put(0,100){\line(1,0){450}}
\put(382,0){\line(0,1){200}}
\put(32,96){\line(0,1){8}}
\put(176,96){\line(0,1){8}}
\put(317,96){\line(0,1){8}}
\put(425,96){\line(0,1){8}}
\put(384,185){$y$}
\put(21,108){$x_4$}
\put(176,108){$x_3$}
\put(311,108){$x_2$}
\put(425,108){$x_1$}
\put(445,90){$x$}
\end{picture}
\caption{\label{fig:gofx}The graph of quartic function $y=G(x)$}
\end{figure}

Let us label the roots of the quartic equation $G(x)=0$ as $x_i$ in descending
order (we are considering the case when we have four real roots, see Fig.~\ref{fig:gofx})
$x_4<x_3<x_2<0<x_1$. We shall restrict attention to the following ranges for
the coordinates.
\beq
x&\in&[x_2,x_1]\\
y&\in&[x_3,x_2]\\
\phi&\in&[\,0,2\pi)\\
t&\in&(-\infty,\infty)
\eq
where $\phi$ is defined by
\be
\phi=\sh{G'(x_2)}\alpha.
\ee
The range of $\phi$ has been chosen to eliminate the cosmic string on one half of the axis.

\begin{figure}[ht]
{\leavevmode\epsfxsize \hsize\epsfysize 10cm\epsfbox{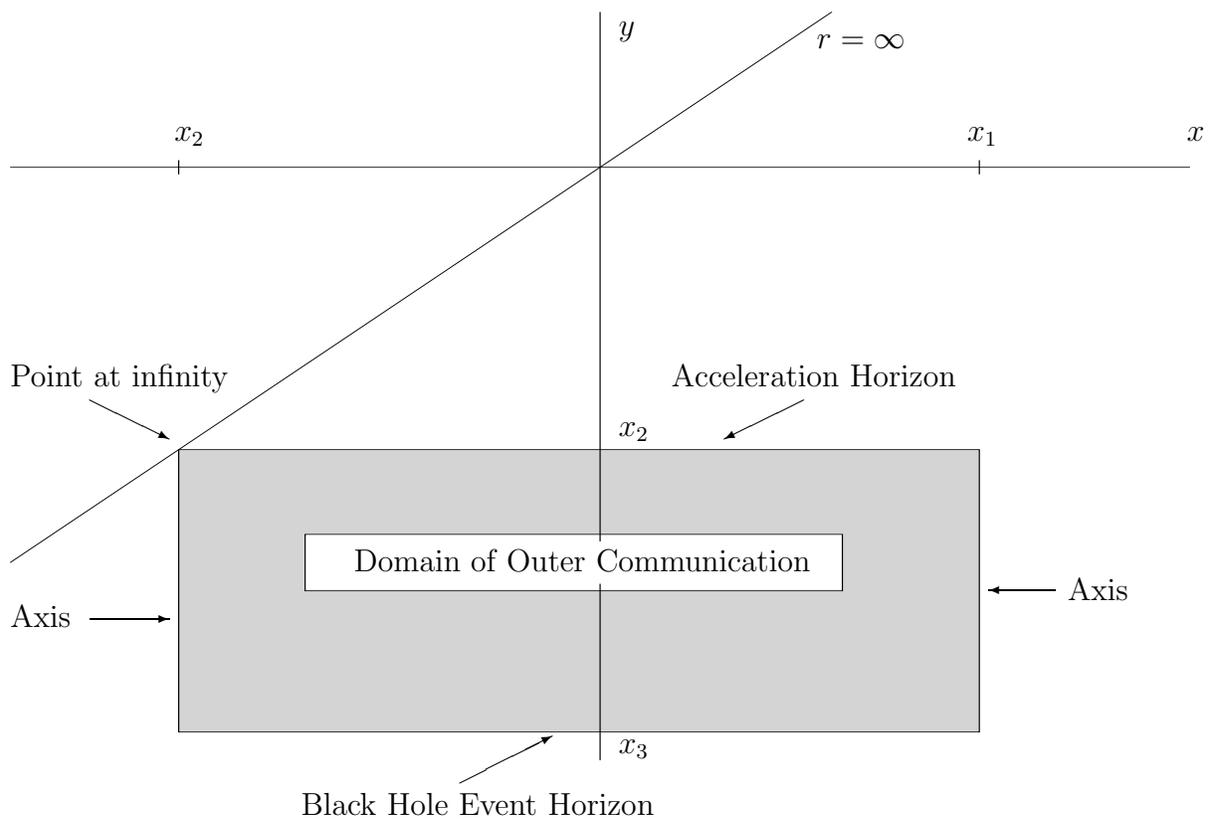}}
\begin{picture}(100,-600)
\put(130,92){Domain of Outer Communication}
\put(62,255){$x_2$}
\put(362,255){$x_1$}
\put(230,295){$y$}
\put(445,255){$x$}
\put(230,23){$x_3$}
\put(230,143){$x_2$}
\put(0,162){Point at infinity}
\put(250,162){Acceleration Horizon}
\put(110,0){Black Hole Event Horizon}
\put(0,71){Axis}\put(400,81){Axis}
\put(305,290){$r=\infty$}
\put(30,157){\vector(2,-1){30}}
\put(30,74){\vector(1,0){30}}
\put(395,85){\vector(-1,0){25}}
\put(300,157){\vector(-2,-1){30}}
\put(170,12){\vector(2,1){35}}
\end{picture}
 \caption{\label{fig:doc}The Horizon Structure of the $C$-metric}
\end{figure}

This means that $0<r<\infty$, the singularity as $r\rightarrow0$ corresponds to
$y\rightarrow-\infty$ whilst $r\rightarrow\infty$ corresponds to the point
$x=x_2,\ y=x_2$. There are two horizons that interest us: 
a black hole event
horizon at $y=x_3$ and an acceleration horizon at $y=x_2$. In addition there
is an inner horizon at $y=x_4$. With these choices we are left with a cosmic strut being
the section of the axis $x=x_1,$ $x_3<y<x_2$. The domain of outer communication is illustrated in Fig.~\ref{fig:doc}.

\begin{figure}[ht]
{\leavevmode\epsfxsize \hsize\epsfysize 10cm\epsfbox{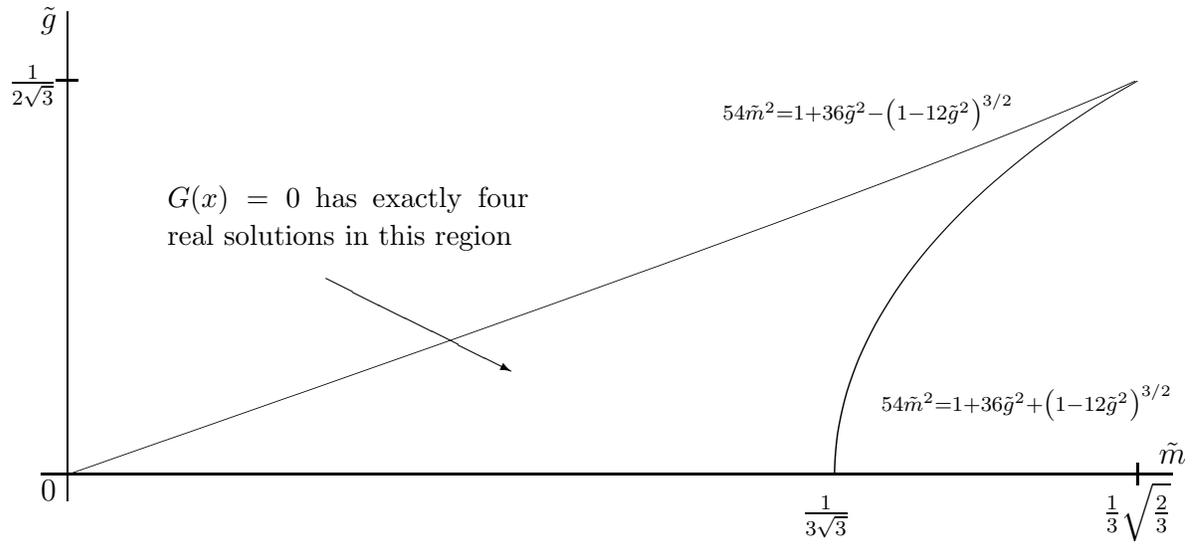}}
\begin{picture}(100,-600)
\put(22,56){\line(1,0){428}}
\put(32,46){\line(0,1){185}}
\put(437,52){\line(0,1){8}}
\put(28,205){\line(1,0){8}}
\put(22,46){$0$}
\put(445,60){$\tilde m$}
\put(22,225){$\tilde g$}
\put(10,201){$\frac1{2\sqrt3}$}
\put(424,38){$\frac13\sqrt\frac23$}
\put(310,38){$\frac1{3\sqrt3}$}
\put(280,190){$\scriptstyle54\tilde m^2=1+36\tilde g^2-\(1-12\tilde g^2\)^{3/2}$}
\put(340,80){$\scriptstyle54\tilde m^2=1+36\tilde g^2+\(1-12\tilde g^2\)^{3/2}$}
\put(70,150){\parbox{48mm}{\sloppy\small $G(x)=0$ has exactly four real solutions in this region}}
\put(130,130){\vector(2,-1){70}}
\end{picture}
\caption{\label{fig:parsp}The Parameter space for $\tilde m, \tilde g$}
 \end{figure}

{}From the metric we may read off the norm of the Killing bivector, $\rho$, for the $C$-metric:
\be
\rho=r^2\sqrt{-G(x)G(y)}.
\ee
We have imposed the condition that $G(x)=0$ have four real roots, this
condition defines a region in the parameter space $(\tilde m,\tilde g)$ shown in Fig.~\ref{fig:parsp}.

It is a simple matter now to determine the Ernst potentials derived from the
angular Killing vector for the $C$-metric. They are presented below:
\beq
\epsilon&=&-r^2G(x)-{\tilde g^2\over A^2}(x-x_2)^2,\\
\psi&=&{\tilde g\over A}(x-x_2).\\
\eq

\subsection{Melvin's Magnetic Universe and The Ernst Solution}
\label{sect:melmag}

In this section we look at the result of performing a Harrison 
transformation
 Eq.~(\ref{eq:ht1}) on Minkowski space and the $C$-metric. We will be applying the
transformation derived from consideration of the angular Killing vector
$\partial/\partial\phi$.

Let us write Minkowski space in terms of cylindrical polar coordinates, thus
\be
\b g=-\ddd{\tilde t}{\tilde t}+\ddd rr+r^2\ddd\phi\phi+\ddd xx.
\ee
The Ernst potentials derived from the angular Killing vector are
\beq
\epsilon&=&-r^2,\\
\psi&=&0.
\eq
Performing the Harrison transformation gives the new Ernst Potentials:
\beq
\epsilon&\mapsto&-r^2\(1+\sq B_0^2r^2\)^{-1},\\
\psi&\mapsto&\frc 12B_0r^2\(1+\sq B_0^2r^2\)^{-1},
\eq
and hence the new metric is
\beq
\b g&=&\Lambda^2(-\ddd{\tilde t}{\tilde t}+\ddd rr+\ddd xx)+r^2\Lambda^{-2}\ddd\phi\phi,
\label{eq:melvin}\\
\Lambda&=&1+\sq B_0^2r^2.
\eq
The electromagnetic field is given by
\be
\b F=
\frc{B_0r\dd r\w\dd\phi}{\(1+\frac14B_0^2r^2\)^2},
\ee
This solution is Melvin's Magnetic Universe \cite{Melvin}.
The Melvin solution represents a uniform tube of magnetic lines of flux in
stable equilibrium with gravity. The transverse magnetic pressure balancing the
attractive gravitational force. 

We now proceed to apply the Harrison transformation to the $C$-metric, the new
Ernst potentials are easily derived:
\beq
\epsilon&=&-\Lambda^{-1}\(r^2G(x)+\frc{\tilde g^2}{A^2}(x-x_2)^2\),\\
\psi&=&\Lambda^{-1}\(\frc{\tilde g}A(x-x_2)+\frc B2\(r^2G(x)
+\frc{\tilde g^2}{A^2}(x-x_2)^2\)\),
\eq
with
\be
\Lambda=\(1+\frc{B\tilde g}{2A}(x-x_2)\)^2+\frc14B^2r^2G(x).
\ee
The metric and electromagnetic field tensor are transformed into (Ernst \cite{Ernst}):
\beq
\b g&=&\Lambda^2r^2\(\frc{\dd x\otimes\dd x}{G(x)}-\frc{\dd y\otimes\dd y}
{G(y)}+G(y)\dd t\otimes\dd t\)+r^2G(x)\Lambda^{-2}\ddd\alpha\alpha,\nonum\\
&&\ \\
\b F&=&\dd\psi\w\dd\alpha.
\eq
 
The great advantage of performing a Harrison Transformation to the $C$-metric
is that it allows us to eliminate the conical singularity from the entire axis.
We do this by carefully choosing the parameter $B$, 
it turns out that the condition to achieve this is given by:
\be
\left.\(\frc1{\Lambda^2}\frc{dG(x)}{dx}\)\right|_{x=x_2}\!\!\!\!\!\!\!\!+
\left.\(\frc1{\Lambda^2}\frc{dG(x)}{dx}\)\right|_{x=x_1}\!\!\!\!\!\!\!\!=0.
\label{eq:nonodal}
\ee
In the limit $mA,gA,gB\ll 1$ this equation reduces to Newton's Second Law,
\be
gB=mA
\ee
The Ernst Solution represents a black hole monopole undergoing a
uniform acceleration due to the presence of a cosmological magnetic field. This
solution has an electric counterpart, obtained by performing a duality
transformation to the solution. However instanton considerations concentrate on the magnetic situation, as electron-positron pair creation will rapidly destroy any electric field capable of producing electrically charged black hole pairs.
 
\subsection{The Ernst Solution in terms of Elliptic Functions}
\label{sect:ernst-elliptic}

In this section we will draw on the properties of elliptic functions.
For a brief summary of all the results we will need and in order to establish our conventions, see Appendix~\ref{sect:elliptic}.

\begin{figure}
{\leavevmode\epsfxsize \hsize\epsfysize 8cm\epsfbox{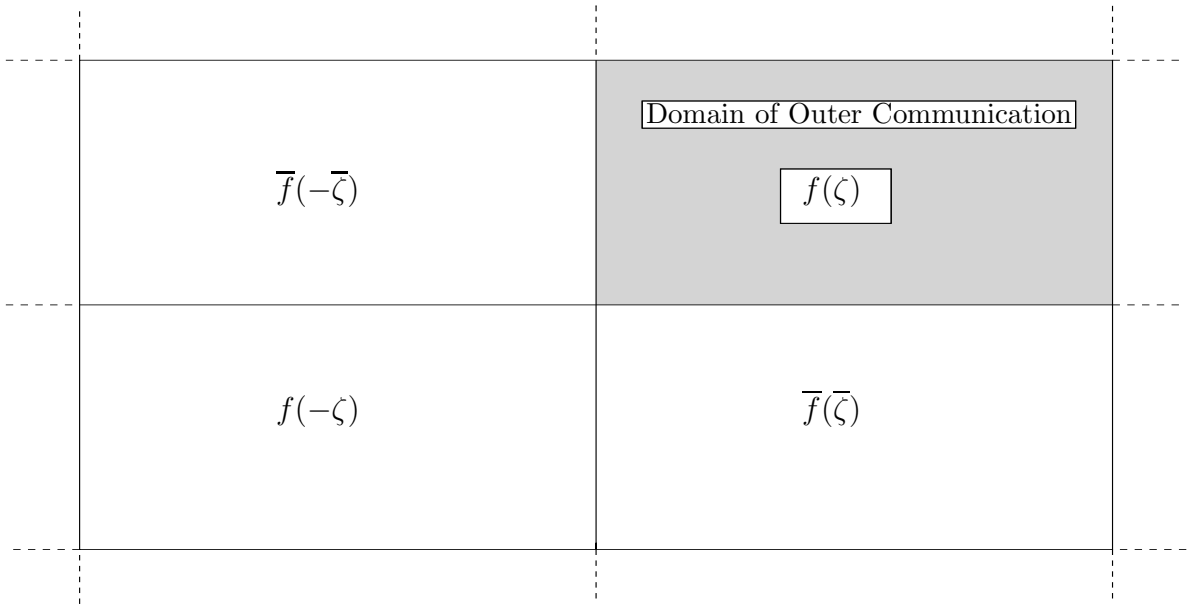}}
\hspace{-20pt}\begin{picture}(0,0)(-92,-121)
\put(150,82){\small Domain of Outer Communication}
\put(209,53){$f(\zeta)$}
\put(209,-30){$\overline f(\overline\zeta)$}
\put(10,53){$\overline f(-\overline\zeta)$}
\put(10,-30){$f(-\zeta)$}
\end{picture}
\caption{\label{fig:sch}Using Schwarz reflection to extend the analytic function $f$.}
\end{figure}

As we remarked previously the norm of the Killing bivector, $\rho$ for the
$C$-metric (and Ernst solution) is simply given by:
\be
\rho=r^2\sqrt{-G(x)G(y)}\ .
\ee
 The induced metric on the two-dimensional space of orbits of the group action generated by the 
the Killing vectors, $\mii$, has the form
\be
\gii=\frc{\ddd xx}{G(x)}-\frc{\ddd yy}{G(y)}\ .
\ee
We may calculate $z$, the harmonic conjugate to $\rho$, from the Cauchy-Riemann equations:
\be
\sqrt{G(x)}\pp\rho x=\sqrt{-G(y)}\pp z y\ ,\qquad\quad
\sqrt{-G(y)}\pp\rho y=-\sqrt{G(x)}\pp z x\ .
\ee
This leads to
\be
z=\h r^2(G(x)+G(y))+\h g^2(x+y)^2+{m\over A}(x+y)+{\rm
constant.}
\ee
We shall denote by $z_A$, $z_N$ and $z_S$ the images of the acceleration horizon,  and the north and 
south poles of event horizon. It is also useful to define
\be
k^2=\frc{z_N-z_S}{z_A-z_S}=\frc{\(x_1-x_2\)\(x_3-x_4\)}{\(x_1-x_3\)\(x_2-x_4\)}\ .
\label{eq:defmod}
\ee
The quantity $k$ turns out to be the modulus of many of the elliptic 
functions we shall be using.
We will transform coordinates so that
\be
\frc\chi M=\int_{x_2}^x\frc{dt}{\sqrt{G(t)}}\ ,\qquad\mbox{and}
\quad\frc\eta M=\int_y^{x_2}\frc{dt}{\sqrt{-G(t)}}\ .
\ee
The value of $M$ is given by
\be
M^{2}=e_1-e_3
\ee
where $e_i=\wp(\omega_i)$, the Weierstrass $\wp$-function being formed with 
the invariants $g_2$ and $g_3$ of $G$ given by
\beq
g_2&=&\frc{1-12\tilde g^2}{12}\ ,\\
g_3&=&\frc{1+36\tilde g^2-54\tilde m^2}{216}\ .
\eq
Letting $\zeta=\chi+i\eta$ we notice that $f(\zeta)=z(\zeta)-i\rho(\zeta)$ is an analytic function defined on $\mii$, the
(two-dimensional section of) the domain of outer communication. In these coordinates the domain of outer communication a rectangle in the
complex $\zeta$-plane. We now use Schwarz reflection in the boundaries (where
$\rho=0$), see Fig.~\ref{fig:sch}, to analytically extend this function to all values of $\zeta$.

On each rectangle $f(\zeta)$ takes the value indicated (and by the
permanence of functional relations under analytic continuation they apply
everywhere). We may proceed to reflect in the new boundaries, what we find is
that $f(\zeta)=f(\zeta+2K)$ and $f(\zeta)=f(\zeta+2iK')$ where
\be
\frc KM=\int_{x_2}^{x_1}\frc{dt}{\sqrt{G(t)}}\quad\mbox{and}\quad
\frc{K'}M=\int_{x_3}^{x_2}\frc{dt}{\sqrt{-G(t)}}
\ee
and hence $f$ is an even doubly periodic meromorphic function, i.e., a map
between two compact Riemann surfaces, namely a torus, $T$ and the Riemann
sphere ${\Bbb C}^\infty$.

Applying the Valency theorem, we deduce that $f$ is exactly $n$-1
for some $n$ (and $n\geq2$ as the sphere and torus are not homeomorphic). We
can find $n$ by examining the pre-image of infinity, $f^{-1}\{\infty\}=\{0\}$.
We must work out the multiplicity, a simple calculation tells us:
\be
f(\zeta)=\frc{2L^2}{\zeta^2}+O(1)
\ee
with $M=AL$. There is a second order pole at $\zeta=0$. Therefore $f:T\rightarrow{\Bbb C}^\infty$ is
exactly 2-1. Clearly $f$ restricted to $\mii$,
$f|_{\cal M_{\rm II}}:\mii\rightarrow\{z-i\rho|\rho>0\}$ is 1-1.
 
As the map $f$ is a doubly periodic even meromorphic
function, another application of the Valency theorem shows that any analytic
map
$:T\rightarrow{\Bbb C}^\infty$ can be expressed in terms of the Weierstrass
$\wp$-function and its derivative. Our map is especially simple
\beq
f(\zeta)&=&2L^2(\wp_\Omega(\zeta)+\alpha),\qquad\alpha
\mbox{ some real constant,}
\label{eq:fwp}\\
\Omega&=&2K{\Bbb Z}+2iK'{\Bbb Z}.
\eq
Without loss of generality we set $\alpha=0$. The critical points of $\wp_\Omega(\zeta)$ are the four corners of $\mii$ where
$\wp_\Omega'(0)=\infty$ and $\wp_\Omega'(K)=\wp_\Omega'(iK')=
\wp_\Omega'(K+iK')=0$, this follows from the observation that the
mapping fails to be conformal at these points, or alternatively
by noticing it as a
particular property of the $\wp$-function. We remark that $\wp_\Omega'(\zeta)$
is exactly 3-1 and we have three points where $\wp_\Omega'(\zeta)=0$ and three
(coincident) points where $\wp_\Omega'(\zeta)=\infty$. Hence we have found all
the critical points of the map. Except at the critical points, the function 
$f|_{\mii}$ is invertible and we conclude that $(\rho,z)$ provide a coordinate system for the
domain.

We now write the solution in terms of the coordinates $(\chi,\eta)$, 
after defining $\kappa=G'(x_2)$ and
\be
q=\frc{\kappa\tilde g}{AM^2}=\frc{\kappa\tilde gL}{M^3}
\ee
we have
\be
\sqrt{G(x)}=\frc{-G'(x_2)\wp'(\chi/M)}{4\(\wp(\chi/M)-\frac1{24}G''(x_2)\)^2}
=\frc{\kappa\sn\chi\cn\chi\dn\chi}{2M\(\cn^2\chi+D\sn^2\chi\)^2}
\ee
\be
x-x_2=\frc{G'(x_2)}{4\(\wp(\chi/M)-\frac1{24}G''(x_2)\)}=\frc{\kappa\sn^2\chi}{4M^2\(\cn^2\chi+D\sn^2\chi\)}
\ee
and
\be
\sqrt{-G(y)}=\frc{\kappa\sn\eta\cn\eta\dn\eta}{2M\(1-D\sn^2\eta\)^2}
\ee
\be
y-x_2=\frc{-\kappa\sn^2\eta}{4M^2\(1-D\sn^2\eta\)}
\ee
for the constant $D=(1+k'^2)/3-G''(x_2)/24M^2$.

The metric takes the form
\be
\b g=-V\ddd tt+X\ddd\phi\phi+\Sigma\(\ddd\chi\chi+\ddd\eta\eta\)
\ee
where
\be
X=\frc{4L^2(1-D\sn^2\eta)^2\sn^2\chi\cn^2\chi\dn^2\chi}{\Lambda^2\(\cn^2\chi+D\sn^2\chi\)^2\(\sn^2\chi+\sn^2\eta\cn^2\chi\)^2}
\ee
\be
V=\frc{4\Lambda^2L^2\(\cn^2\chi+D\sn^2\chi\)^2\sn^2\eta\cn^2\eta\dn^2\eta}{\(\sn^2\chi+\sn^2\eta\cn^2\chi\)^2\(1-D\sn^2\eta\)^2}
\ee
\be
\Sigma=\frc{16\Lambda^2L^2\(\cn^2\eta+D\sn^2\eta\)^2(1-D\sn^2\eta)^2}{\kappa^2\(\sn^2\chi+\sn^2\eta\cn^2\chi\)^2}
\ee
\be
\Lambda=\(1+\frc{B_0q\sn^2\chi}{8(\cn^2\chi+D\sn^2\chi)}\)^2+
\frc{B_0{}^2L^2\(1-D\sn^2\eta\)^2\sn^2\chi\cn^2\chi\dn^2\chi}
{\(\cn^2\chi+D\sn^2\chi\)^2\(\sn^2\chi+\sn^2\eta\cn^2\chi\)^2}
\ee
\be
\rho=\frc{4L^2\sn\chi\cn\chi\dn\chi\sn\eta\cn\eta\dn\eta}{\(\sn^2\chi+\sn^2\eta\cn^2\chi\)^2}
\ee
and
\be
z-i\rho=2L^2\wp(\chi+i\eta).
\ee
We have written $B_0$ for the Harrison transformation parameter above and reserve
$B$ for the magnetic potential.
The $\wp$-function is with respect to the lattice $2K{\Bbb Z}+2iK'{\Bbb Z}$, 
so that $2L^2=z_A-z_S$.
In addition the magnetic field is given by
\be
\b F=\dd E\w\dd\phi
\ee
with
\beq
E&=&\frc1\Lambda\(\frc{q\sn^2\chi}{4(\cn^2\chi+D\sn^2\chi)}+\frc{2B_0L^2\(1-D\sn^2\eta\)^2\sn^2\chi\cn^2\chi\dn^2\chi}{\(\cn^2\chi+D\sn^2\chi\)^2\(\sn^2\chi+\sn^2\eta\cn^2\chi\)^2}\right.\nonum\\
{}&&\hspace{2cm}\left.
+\frc{B_0q^2\sn^4\chi}{32\(\cn^2\chi+D\sn^2\chi\)^2}\).
\eq
We will need to investigate the behaviour of $X$, $E$ and $\rho$ near the axis $\chi=0$, we find
\be
X=O(\chi^2)
\ee
\be
E=O(\chi^2)
\ee
and
\be
\rho=\frc{4L^2\cn\eta\dn\eta}{\sn^3\eta}\,\chi+O\(\chi^3\).
\ee
Near the other axis $\chi=K$ we discover, setting $u=K-\chi$,
\be
X=O(u^2),
\ee
\beq
E&=&\frc2{B_0+8D/q}+O(u^2) 
\eq
and
\be
\rho=4L^2k'^2\sn\eta\cn\eta\dn\eta\, u+O\(u^3\).
\ee
Near infinity, setting $\chi=R^{-1/2}\sin\theta$ and 
$\eta=R^{-1/2}\cos\theta$, we have
\be
X=\frc{4}{B_0{}^4L^2\sin^2\theta}\,\frc1R+O\(\frc1{R^2}\)
\ee
and
\be
E=\frc2{B_0}-\frc{2}{B_0{}^3L^2\sin^2\theta}\,\frc1R+O\(\frc1{R^2}\).
\ee
Finally we find that $\rho$ behaves as
\be
\rho={4L^2\sin\theta\cos\theta}R +O\(\frc1R\).
\ee

\subsection {The $C$-metric in terms of Elliptic Functions}
\label{sect:C-elliptic}

As a special case of the previous section we set the cosmological magnetic field $B_0$ to zero. Doing so leads to a different behaviour near infinity. We find that
\beq
X&=&O\(\chi^2\)\\
E&=&O\(\chi^2\)
\eq
close to the axis $\chi=0$. Near the other axis $u=0$ with $u=K-\chi$,
\beq
X&=&O\(u^2\)\\
E&=&\frc q{4D}+O\(u^2\)
\eq
whilst near infinity the behaviour is quite different from that of the Ernst Solution, and we have
\beq
X&=&4L^2\sin^2\theta\,R+O(1)\\
E&=&\frc q{4\sin^2\theta}\,\frc1R+O\(\frc1{R^2}\)
\eq
We will therefore need to impose different boundary conditions to prove 
the uniqueness of this solution. This will be done in
Subsect.~\ref{sect:Cboundary}.

\section {Determination of the Parameters of the Ernst Solution}
\label{sect:parameters}

We now present a couple of technical lemmas that will enable us to determine 
the parameters $\tilde m$ and $\tilde g$ 
from an Ernst solution by looking closely at its behaviour on the axis and as
 one goes off towards infinity. This will be important for our
 discussion of the uniqueness theorems in the next sections.  
Given a candidate spacetime we need to find an Ernst solution that 
coincides asymptotically (and to the right order) on the axis and off towards 
infinity. In addition to complete the uniqueness result we need to have both 
solutions defined on a common domain. This means that the quantity $k$ 
defined by Eq.~(\ref{eq:defmod}) must be the same for each solution.
If we can find such an Ernst solution then we may use the 
divergence identity from Sect.~\ref{sect:divergence} to prove the uniqueness of the Ernst solution. 

The boundary conditions we will need determine $B_0$ directly. The quantities 
$L$ and $q/D$ may be regarded as given. 
In addition, as we have just remarked we may assume knowledge of 
$k$ the modulus of the elliptic functions.

We break the proof into two lemmas. Firstly we prove that the 
parameters $D$ and $k$ uniquely determine $\tilde m$ and $\tilde g$. 
 
\noindent{\bf Lemma. }{\em Given the modulus $k\in(0,1)$ and $D\in[0,k']$, where $k'$ is the complementary modulus
there exist values of $\tilde m$ and $\tilde g$ such that
\be
g(\chi)=\frc{\kappa\sc\cc\dc}{2M(\cn^2\chi+D\sn^2\chi)^2}=\sqrt{G(x)}=\sqrt{1-x^2-2\tilde mx^3-\tilde g^2x^4}
\ee
where
\be
\frc{dx}{\sqrt {G(x)}}=\frc{d\chi}M,
\ee
the values of $M$ and $\kappa$ being determined from $D$ and $k$.}

\noindent{\em Proof:} To begin with we investigate the turning points of $g(\chi)$ for $\chi\in[K/2,K]$. We therefore differentiate:
\beq
\frc {2M}\kappa\,\frc{dg(\chi)}{d\chi}&=&\frc{\cn^2\chi\dn^2\chi}{(\cn^2\chi+D\sn^2\chi)^2}
-\frc{\sn^2\chi\dn^2\chi}{(\cn^2\chi+D\sn^2\chi)^2}\nonum\\
&&{}-\frc{k^2\sn^2\chi\cn^2\chi}{(\cn^2\chi+D\sn^2\chi)^2}
-\frc{4(D-1)\sn^2\chi\cn^2\chi\dn^2\chi}{(\cn^2\chi+D\sn^2\chi)^3}\ .
\eq
Setting this equal to zero we find that
\be
D=\frc{3\sn^2\cz\cn^2\cz\dn^2\cz+\cn^4\cz\dn^2\cz-k^2\sn^2\cz\cn^4\cz}
{3\sn^2\cz\cn^2\cz\dn^2\cz+\sn^4\cz\dn^2\cz+k^2\sn^4\cz\cn^2\cz}\ .
\ee
We shall now prove that $D\in[0,k']$ is in one to one correspondence with the values $\cz\in[K/2,K]$. On this region $\sn^2\cz$ varies monotonically from $1/(1+k')$ to unity.
We make the substitutions:
\beq
\sn^2\cz&=&S\\
\cn^2\cz&=&1-S\\
\dn^2\cz&=&1-k^2S
\eq
Note that when $S=1/(1+k')$, $D=k'$ and when $S=1$, we find $D=0$. We now prove that $D(S)$ is monotonic decreasing:
\be
\frc{dD}{dS}=-\frc{h(k,S)}
{S^2\left[3-2(1+k^2)S+k^2S^2\right]^2}
\ee
where the function $h(k,S)$ is defined by
\beq
h(k,S)&=&3-4(1+k^2)S+2(2k^4+k^2+2)S^2-4k^2(1+k^2)S^3+3k^4S^4\nonum\\
&=&(1-k^2S^2)^2+2[1-(1+k^2)S+k^2S^2]^2+2(1-k^2)^2S^2\ge0
\eq
with equality if and only if $S=1$ and $k=1$. This establishes the strict monotonicity. 
Hence we may write $\cz=\cz(D)$. Next we calculate $M^2$ by using $G''(0)=-2$, i.e.,
\be
2\left.\frc{d^2\log g(\chi)}{d\chi^2}\right|_{\chi=\cz}\!\!\!\!=-\frc2{M^2}\ .
\ee
This equation can be written (on eliminating $D$) as
\be
\sqrt{\sn\cz\cn\cz\dn\cz}\,\frc{d^2}{d\cz^2}\(\frc1{\sqrt{\sn\cz\cn\cz\dn\cz}}\)=\frc1{2M^2}
\ee
in terms of the variable $S$ introduced earlier we have,
\be
\frc1{M^2}=
\frc{(1-k^2S^2)^2+2[1-(1+k^2)S+k^2S^2]^2+2(1-k^2)^2S^2}
{2S(1-S)(1-k^2S)}
\ee
The function $M^2$ is monotonically decreasing on $S\in[1/(1+k'),1]$ i.e., on $\cz\in[K/2,K]$ with $M^2(1)=0$ and
\be
M^2_{{}_{\it max}}=\frc1{1+k'^2}.
\ee
The derivative with respect to $S$ of $M^2$ is given by
\be
\frc{d\,M^2}{dS}=\frc{6(1-k^2S^2)(k'^2+k^2(1-S)^2)(1-S(1+k'))(1-S(1-k'))}{\((1-k^2S^2)^2+2[1-(1+k^2)S+k^2S^2]^2+2(1-k^2)^2S^2\)^2}\ge0.
\ee
We have equality only when $\cz=K/2$.

Having found $\cz$ and $M^2$ we may read off $\kappa$ by noting that $G(0)=1$, thus
\be
\kappa=\frc{2M(\cn^2\cz+D\sn^2\cz)^2}{\sn\cz\cn\cz\dn\cz}.
\ee
We may now go on to find the value of $\tilde g$. We use the relation (\ref{eq:esquared}) that
\beq
1-12\tilde g^2&=&\sh
M^4\left[(\epsilon_1-\epsilon_2)^2+(\epsilon_2-\epsilon_3)^2+(\epsilon_3-\epsilon_1)^2\right]\nonum\\
&=&M^4(1-k^2+k^4)\nonum\\
&=&M^4(1-k'^2+k'^4).\label{eq:findg}
\eq
We have already seen that $M^2\le1/(1+k'^2)$. Hence the RHS of Eq.~(\ref{eq:findg}) is bounded above by
\be
1-\frc{3k'^2}{(1+k'^2)^2}\le1.
\ee
That is to say the value $\tilde g$ is uniquely determined.
We now make use of the discriminant expression (\ref{eq:ecubed}) to write
\be
\frc{(\epsilon_1-\epsilon_2)^2(\epsilon_2-\epsilon_3)^2(\epsilon_3-\epsilon_1)^2}{\((\epsilon_1-\epsilon_2)^2+(\epsilon_2-\epsilon_3)^2+(\epsilon_3-\epsilon_1)^2\)^3}=\frc{g_2{}^3-27g_3{}^2}{54g_2{}^3}
\ee
i.e.,
\be
\frc{54k^4k'^4}{(1+k^4+k'^4)^3}=\frc{(1-12\tilde g^2)^3-(1+36\tilde g^2-54\tilde m^2)^2}{(1-12\tilde g^2)^3}.
\ee
This determines $\tilde m$. Observe that the LHS takes values between $[0,1]$ attaining its upper bound only when $k^2=1/2$. We take
\be
54\tilde m^2=1+36\tilde g^2-\left[1-\frc{54k^4k'^4}{(1+k^4+k'^4)^3}\right]^{1/2}(1-12\tilde g^2)^{3/2}
\label{eq:m21}\ee
for $k^2\le1/2$ and
\be
54\tilde m^2=1+36\tilde g^2+\left[1-\frc{54k^4k'^4}{(1+k^4+k'^4)^3}\right]^{1/2}(1-12\tilde g^2)^{3/2}
\label{eq:m22}\ee
when $k^2\ge1/2$. This is because when $k\rightarrow0$ our solutions lie on the line given by Eq.~(\ref{eq:m21}) 
while when $k\rightarrow1$ they satisfy Eq.~(\ref{eq:m22}), see Fig.~\ref{fig:parsp}. Continuity then determines which solution 
to take as we increase $k$ from zero to one. $\square$

Thus it suffices to find $D$ from the quantities directly read off from the boundary conditions.
 These being $L$ and $q/D$. For the next Lemma it is useful to define
\be
\Delta=\frc q{DL}=\frc{\kappa\tilde g}{M^3D}
\ee
which we may assume is given.

\noindent{\bf Lemma. }{\em Given the modulus $k$ and the quantity $\Delta\!^2$ defined by
\be
\Delta\!^2(D)=\frc{\kappa^2\tilde g^2}{M^6D^2},
\ee
we can invert to give $D=D(\Delta\!^2)$ provided 
\be
\Delta\!^2\ge\frc{16(1+3k'^2+k'^4)}{(1+k')^2}\, .
\ee}

\noindent{\em Proof:}  Firstly we show that
\be
\Delta\!^2(k')=\frc{16(1+3k'^2+k'^4)}{(1+k')^2}
\ee
rising monotonically to infinity. To see that $\Delta\!^2$ is increasing on $S\in[1/(1+k'),1]$,
 we examine its derivative. We find
\be
\frc{d\Delta\!^2}{dS}=\frc{64(1-(1-k')S)((1+k')S-1)f(k,S)h(k,S)}{S^2(1-S)^2\(1-k^2S^2+2k'^2S\)^3 \(1-S^2+k'^2+\(1+k^2\)(1-S)^2\)^3}
\ee
where we have defined
\beq
f(k,S)=&&
-3
+\(16-20k^2\)S
+\(36-36k^2+28k^4\)S^2
\nonum\\&&{}
+\(8+48k^2+44k^4-8k^6\)S^3
+\(-4-12k^2-50k^4-44k^6-4k^8\)S^4
\nonum\\&&{}
+\(8k^2+36k^6+24k^8\)S^5
+\(-4k^4+4k^6-20k^8\)S^6
+4k^8S^7
-k^8S^8.\nonum\\&&\ 
\eq
We may write $f(k,S)$ in an explicitly non-negative form for $S,k\in[0,1]$, 
\beq
f(k,S)=&&
1
+2\(1-S+5k'^2S\)^2
+2\(2+11k^2+11k'^2k^2\)S^2
\nonum\\&&{}
+4\(18k^2+5k^4+2k'^6\)S^3(1-S)^5\nonum\\&&{}
+\(4k^2k'^6+24k'^4+12k'^2+256k^2k'^2\right.
\nonum\\&&{}\hspace{4cm}\left.+96k^4k'^2+28k^2+318k^4\)S^4(1-S)^4
\nonum\\&&{}
+\(440k^2k'^2+220k^4k'^2+8k^8+460k^4+64\)S^5(1-S)^3\nonum\\&&{}
+\(28+28k'^2+560k'^2k^2+232k'^2k^4+364k^4+28k^8\)S^6(1-S)^2\nonum\\&&{} +\(24+216k'^2k^2+100k^4k'^2+128k^4+20k^8\)S^7(1-S)\nonum\\&&{}
+\(1+3k'^8+28k^2k'^2+28k^2\)S^8\ge0.
\eq
Thus we have proved that the derivative of $\Delta\!^2$ is non-negative on the required 
domain and as it is clearly non-constant the derivative has isolated zeros (being analytic in $S$),
 therefore we may conclude that $k$ and the value of $\Delta\!^2$ in the range $[\Delta\!^2(k'),\infty)$
 uniquely determine the mass and charge parameters, $\tilde m$ and $\tilde g$ for a
 suitable Ernst solution. Having done so we may then construct $M$ which in turn 
determines the acceleration from the relation $A=M/L$. Thus we have one 
constraint on the range of the parameters representing the Ernst solution 
when we write it in terms of the elliptic functions introduced, namely
\be
\Delta\!^2\ge\frc{16\(1+3k'^2+k'^4\)}{\(1+k'\)^2}.
\ee
It remains an open question whether there exists other solutions to the
 Einstein-Maxwell system that behave asymptotically like the Ernst solutions
 that violate this condition which have no naked singularities or other serious defects.

\section{Black Hole Uniqueness Theorems for The Ernst Solution and $C$-metric}
\label{sect:unique}

The uniqueness theorems we will be presenting for the $C$-metric and Ernst solutions are schematically identical to the proof of the uniqueness theorem for the Kerr-Newman black hole. However, the devil is in the details. The most difficult complication arises because of the presence of another horizon: the acceleration horizon. The boundary conditions are then given on five distinct regions: two horizons, two sections of the axis and at infinity. Infinity will
be represented as a single point on the boundary after a suitable transformation of coordinates. The other portions of the boundary form a rectangle in these coordinates. The fact that not all rectangles are conformally homeomorphic will be the major complicating factor. Contrast this situation with what happens in the Kerr-Newman uniqueness theorem. In this case there are four parts to the boundary, this is represented by a semi-infinite rectangle, the non-existent fourth side mathematically describes arbitrarily large distances.  By a simple scale and an appropriate translation any two such rectangles may be made to coincide.

The uniqueness theorems work by comparing two solutions defined on the same domain, and this is why it is important that the two domains should be conformally homeomorphic, one then uses the divergence result from Sect.~\ref{sect:divergence} to prove uniqueness.

We will be making use of the theory of Riemann surfaces in our deliberations. Riemann surface theory is a valuable asset when it comes to
investigating the introduction of Weyl coordinates, a necessary step in the theorem. We have already seen in the last few sections how effective the application of Riemann surface theory and elliptic function theory was to the
description of the $C$-metric and Ernst solution. We will be making use of these
functions once again when it comes to presenting the appropriate boundary conditions to cause the vanishing of the boundary integral arising from applying
Stokes' theorem to the divergence result.

One might wonder whether it is really necessary to use the elliptic functions. 
 Might there be some benefit from using the $(x,y)$ coordinates that we used when we first wrote down the $C$-metric? The problem with this idea is that for a candidate spacetime we would need to determine $\tilde m$ and $\tilde g$ before we could expand the solution near the axis or near infinity. However, the problem of determining these quantities is extremely difficult, especially when one hasn't yet constructed the $x$ and $y$ coordinates. This is why we really do need to introduce the elliptic functions at an early stage.

\subsection{Hypotheses}
\label{sect:hyp}

In this section we set down in detail the hypotheses will be assuming in order to
establish our uniqueness result. The main differences with the Kerr-Newman
black hole uniqueness theorem occur in the boundary conditions and the
overall horizon structure we will be assuming. It turns out that the different
horizon structure makes proving a uniqueness result much more difficult and
necessitates the use of elliptic functions and integrals.

Below we present the hypotheses we will be using for the rest of this chapter:

\begin{itemize}
\item{Axisymmetry: There exists a Killing vector $m$ such that $\lm\b g=0$, and $\lm\b F=0$ which generates a one-parameter
group of isometries whose orbits are closed spacelike curves.}

\item{Stationarity: There exists a Killing vector $K$ such that $\lk\b g=0$, and $\lk\b F=0$ which generates a one-parameter
group of isometries which acts freely and whose orbits near infinity
are timelike curves.}

\item{Commutivity: $[K,m]=0$.}

\item{Source-free Maxwell equations $\dd\b F=0$ and $\bd\b F=0$
together with the Einstein equations $R_{ab}=8\pi T_{ab}$ where
\be
T_{ab}=\frac1{4\pi}\(F_{ac}F_b{}^c-\frac14g_{ab}F_{cd}F^{cd}\).\nonum
\ee}

\item{The domain of outer communication is connected and
simply-connected.}

\item{The solution has the same horizon structure as the Ernst Solution (and $C$-metric).}

\item{For the Ernst solution uniqueness result we assume the solution 
is asymptotically Melvin's Magnetic Universe, whereas we assume asymptotic
flatness when we come to prove the uniqueness of the $C$-metric.}

\item{Boundary conditions (See Sect.~\ref{sect:bcs}).}

\end{itemize}

\subsection{The Generalized Papapetrou Theorem}
\label{sect:gpap}

Following Carter \cite{Cargese,B-Holes}, we shall prove that there exist
coordinates $t$ and $\phi$ defined globally on the domain of outer communication
so that the metric takes a diagonal form with the Killing vectors $K=\partial
/\partial t$ and $m=\partial/\partial\phi$. In addition we will prove that
the electromagnetic field tensor can be derived from a vector potential
satisfying the appropriate circularity and invariance conditions.

Our starting point is to make the remark that for a Killing vector $K$, 
the Laplacian $-(\bd\dd+\dd\bd)$ acting on $\k$, 
reduces to $\bd\dd\k=2\,\b R(\k)$. Here $\b R(\k) = R_{ab}K^a\e b$ is the 
Ricci form with respect to $\k$. We calculate
\beq
\bd(\k\w\dd \k)&=&-\lk(\dd\k)- \k\w\bd\dd\k\nonum\\
                   &=&-2\,\k\w\b R(\k)\, .
\eq
Hence
\beq
\bd(\k\w\m\w\dd\k)&=&\lm(\k\w\dd\k)+\m\w\bd(\k\w\dd\k)\nonum\\
 &=&2\,\k\w\m\w\b R(\k)\, .
\label{eq:div}
\eq
Now the energy-momentum 1-form of the electromagnetic field with respect 
to $\k$ is given by the formula:
\be
\b T(\k)=\frac1{4\pi}\(\hd(i_K\b F\w\hd\b F)-\frac12\hd(\b F\w\hd\b F)\k\)\, .
\ee
Let us now calculate $\k\w\m\w\b F$. Using $\lk\b F=\lm\b F=0$ and $\bd\b F=0$ we have
\beq
\bd(\k\w\m\w\b F)&=&-\lk(\m\w\b F)-\k\w\bd(\m\w\b F)\nonum\\
&=&\k\w\lm\b F+\k\w\m\w\bd \b F=0.
\eq
Hence $\k\w\m\w\b F=c\b\eta$, for some constant $c$ and $\b\eta$ the volume form. The boundary condition that $\m\rightarrow0$ as one approaches 
the axis requires $c=0$, thus proving
\be
\k\w\m\w\b F=0. 
\label{eq:fcirc}
\ee
Another way to express this is $i_K i_m\hd\b F=0$.
We will also need to examine the analogous quantity $i_K i_m\b F$. We have
\be
\dd i_K i_m\b F=\lk i_m \b F-i_K\lm\b F+i_K i_m\dd\b F=0.
\ee
Using the axis-boundary condition again we see that
\be
i_K i_m \b F=0.
\label{eq:fkm}
\ee
Now use the fact that $\k\w\m\w\b T(\k)=-\hd i_K i_m \hd\b T(\k)$, but
\beq
4\pi\, i_m i_K\hd\b T(\k)&=&i_m(-i_K\b F\w i_K\hd\b F)\nonum\\
&=&i_K i_m\b F\w i_K\hd\b F-i_K\b F\w i_K i_m\hd\b F=0.
\eq
That is to say $\k\w\m\w\b T(\k)=0$.
Einstein's equation, then proves from Eq.~(\ref{eq:div}) that 
$\k\w\m\w\dd\k=c_k\b\eta$ and
$\k\w\m\w\dd\m=c_m\b\eta$. The constants $c_k$ and $c_m$ are then seen be 
zero by another application
 of the boundary condition for $\m$ on the axis.
Let us proceed to define 1-forms $\b\alpha$ and $\b\beta$ by the following:
\beq
\b\alpha&=&i_m\left(\frac{\k\w\m}{\rho^2}\right)\nonum\\
\b\beta&=&-i_K\left(\frac{\k\w\m}{\rho^2}\right)\nonum\\
\eq
where
\be
\rho^2=i_Ki_m(\k\w\m).
\ee
As before, $\rho$ is the norm of the Killing bivector. Notice that 
by construction we have
\beq
\lk\b\alpha&=\lm\b\alpha=0,\nonum\\
\lk\b\beta&=\lm\b\beta=0,
\eq
and also that
\beq
i_K\b\alpha&=1,\qquad i_m\b\alpha&=0,\nonum\\
i_K\b\beta&=0,\qquad i_m\b\beta&=1.
\eq
Together these imply
\be
i_K\dd\b\alpha=i_m\dd\b\alpha=i_K\dd\b\beta=i_m\dd\b\beta=0.
\ee
The integrability conditions we have established may be rewritten as
\be
\k\w\m\w\dd\b\alpha=0\qquad\mbox{and}\qquad\k\w\m\w\dd\b\beta=0.
\ee
Evaluating
\be
i_K i_m (\k\w\m\w\dd\b\alpha) \qquad\mbox{and}\qquad i_K i_m (\k\w\m\w\dd\b\beta)
\ee
we find
\be
\rho^2\dd\b\alpha=0\qquad\mbox{and}\qquad\rho^2\dd\b\beta=0.
\ee
Thus we may write
\be
\b\alpha=\dd t\qquad\mbox{and}\qquad\b\beta=\dd\phi
\ee
in the Domain of Outer Communication, which we have assumed is 
simply-connected. Summarizing, we have shown the existence of coordinates $t$
and $\phi$ satisfying:
\beq
&\k\w\m\w\dd t=0,\qquad i_K\dd t=1,\qquad i_m\dd t=0,\nonum\\
&\k\w\m\w\dd\phi=0,\qquad i_K\dd\phi=0,\qquad i_m\dd\phi=1.
\eq
Turning to the electromagnetic field, Eq.~(\ref{eq:fcirc}) implies that 
$\b F$ takes the form
\be
\b F=\b\alpha\w\b\gamma+\b\beta\w\b\epsilon.
\ee
Making the replacements: 
\beq
\b\gamma&\mapsto&\b\gamma-i_K\b\gamma\,\b\alpha-i_m\b\gamma\,\b\beta,\nonum\\
\b\epsilon&\mapsto&\b\epsilon-i_K\b\epsilon\,\b\alpha-i_m\b\epsilon\,\b\beta,
\eq
we see that $\b F$ changes to 
\be
\b F\mapsto\b F+(-i_m\b\gamma+i_K\b\epsilon)\b\alpha\w\b\beta.
\ee
However, Eq.~(\ref{eq:fkm}) implies
\be
i_K i_m\b F=-i_m\b\gamma+i_K\b\epsilon=0.
\ee
Thus, we may assume without loss of generality that
\beq
&i_K\b\gamma=0,\qquad &i_m\b \gamma=0,\nonum\\
&i_K\b\epsilon=0,\qquad &i_m\b\epsilon=0.
\eq

Maxwell's equation $\dd\b F=0$ and the invariance of $\b F$ under the action 
of the isometries generated by the Killing vectors reduce to the pair of 
equations:
\be
\dd i_K \b F=0,\qquad\mbox{and}\qquad\dd i_m \b F=0.
\ee
Hence we may introduce electrostatic potentials according to
\beq
i_K\b F&=\b \gamma&=-\dd\Phi,\\
i_m\b F&=\b \epsilon&=-\dd\Psi.
\eq
with the potential function $\Phi$ for the electric field and $\Psi$ 
for the magnetic field. It is now a simple matter to define an electromagnetic 
vector potential $\b A$ with $\b F=\dd\b A$ by setting
\be
\b A=\Phi\b\alpha+\Psi\b\beta.
\ee
It is now straightforward to verify that this vector potential satisfies the 
circularity and invariance conditions:
\be
\k\w\m\w\b A=0\qquad\mbox{and}\qquad
\lk\b A=\lm\b A=0. 
\ee

In the next section we will look at how to find a set of coordinates
that cover the entire Domain of Outer Communication. This involves using the
$t$ and $\phi$ coordinates we have just found together with the quantity 
$\rho$ and its harmonic conjugate which shall denote by $z$.

\subsection{Global Coordinates on the Domain of Outer Communication}
\label{sect:globalco}

In contrast to Carter's proof of the uniqueness for the Kerr-Newman black
hole we will be exploiting the theory of Riemann surfaces to justify the
introduction of Weyl coordinates on the Domain of Outer Communication.   
Previously this step in the uniqueness theorems has been done using
Morse theory, however results in Morse theory rely heavily on complex 
variable methods
and one should not be too surprised that the application of Riemann surface
theory can successfully be used to prove the result we need. We
have already seen how useful Riemann surface theory is when we discussed 
the $C$-metric and Ernst
solution in Sects.~\ref{sect:ernst-elliptic} and \ref{sect:C-elliptic}.

There is a natural induced two-dimensional metric on the space of orbits 
of the two-parameter
isometry group generated by the Killing vectors $K$ and $m$. Define $\mii$ as
the space of generic orbits (i.e., two-dimensional orbits) of the isometry 
group
acting on the Domain of Outer Communication. We remark that the fixed point 
set
of the isometry group generated by $\partial/\partial\phi$ is a closed subset 
of the
spacetime. We call this set the {\em axis\/}. Notice too that $\mii$ is open,
connected and non-empty. It is contained in the Hausdorff topological space
consisting of all orbits of the isometry group acting on the spacetime. It is
therefore non-compact (a compact subset of a Hausdorff space is closed, 
but if $\mii$
were both open and closed then it must be equal to the entire Hausdorff space,
as the space of all orbits is connected, and therefore the axis would have to 
be empty which
is not the case). Let the induced metric on $\mii$ be written
\be
\gtii=\tilde g_{{}_{\rm II}\alpha\beta}\ddd{x^\alpha}{x^\beta}.
\ee
Since any two-dimensional metric is conformally flat, we can introduce
1-forms ${\b E}^1,{\b E}^2$ such that
\be
\gtii=\Sigma\gii=\Sigma({\b E}^1\otimes{\b E}^1+{\b E}^2\otimes{\b
E}^2)
\ee
and $\Sigma(p)\neq0$ for $p\in\mii$ and where $\gii$ is flat. Take $p$ a base point in $\mii$. Since
$(\mii,\gii)$
is flat and simply connected, its holonomy is trivial, and we may
parallely transport the 1-forms $\b E^1$ and $\b E^2$ to all other
points in $\mii$ using
\be
\dd{\b E}^\alpha=0.
\ee
Now, as the fundamental group $\pi_1(\mii)=\{\openone\}$ we deduce that
there exists scalars $u,v$ such that
\be
\b E^1=\dd u\quad\mbox{and}\quad\b E^2=\dd v.
\ee
Combining $u$ and $v$ into a complex quantity $\zeta=u+iv$ we see that we 
have a
complex-valued function on the manifold $\mii$, which need not be injective.
However if $q\in\mii$
then there is an open neighbourhood $U$ of $q$ such that
$\zeta|_U:U\rightarrow\zeta (U)$ is one to one and hence $\mii$ is a Riemann
surface.

The quantities $u$ and $v$ do not necessarily constitute a coordinate system 
for the space $\mii$ as the map $\zeta$ on $\mii$ fails to be injective in 
general. However we will show that $\rho$ and its harmonic
conjugate are better behaved in this respect.

Let us consider the Einstein field equation for $\rho$. We will assume the metric
has been put in the form
\be
\b g=-V\ddd tt+W(\ddd\phi t+\ddd t\phi)+X\ddd\phi\phi+\gtii.
\ee
Explicitly we have $\rho^2=XV+W^2$. Define
\be
\(h_{AB}\)=\pmatrix{-V&W\cr W&X}\qquad\mbox{and}\qquad
\(h^{AB}\)=\frc1{\rho^2}\pmatrix{-X&W\cr W&V},
\ee
then $\rho^2=-\det \(h_{AB}\)$. The Ricci tensor can be calculate, in particular we find
\be
{}^4\!R_{AB}h^{AB}=-\frc1{2\rho}\D^\alpha\(\rho h^{AB}\D_\alpha h_{AB}\)
=-\frc1\rho\D^2\rho,
\ee
where $A$ and $B$ refer to the $t$ and $\phi$ coordinates whilst the covariant 
derivatives are with respect to the induced metric on the orbit space. Defining
\be
E_\alpha=F_{t\alpha}\qquad\mbox{and}\qquad B_\alpha=F_{\phi\alpha}
\ee
we have
\beq
{}^4\!R_{tt}&=&2\b{E.E}+\sh VF^2,\\
{}^4\!R_{t\phi}&=&2\b{E.B}-\sh WF^2,\\
{}^4\!R_{\phi\phi}&=&2\b{B.B}-\sh XF^2,
\eq
where we have set
\be
F^2=2(-X\b{E.E}+2W\b{E.B}+V\b{B.B})\rho^{-2}.
\ee
Evaluating
\be
-\frc1\rho\D^2\rho=\frc1{\rho^2}\(-{}^4\!R_{tt}X+2\,{}^4\!R_{t\phi}W+{}^4\!R_{\phi\phi}V\)=0.
\ee
So we have shown that $\rho$ is harmonic. Now any harmonic map may
be written as the real or imaginary part of an analytic function, 
we therefore choose to write
\be
f(\zeta)=z(\zeta)-i\rho(\zeta);\qquad\quad f\ {\rm analytic,}
\ee
so that $z(\zeta)$ is determined up to a constant by integrating the
Cauchy-Riemann equations.

We are now in a position to apply the Riemann Mapping Theorem:

\proclaim The Riemann Mapping Theorem. Any simply-connected Riemann Surface is
conformally homeomorphic to either
\begin{enumerate}
\item[(i)] The Riemann Sphere ${\Bbb C}^\infty$,
\item[(ii)] The complex plane ${\Bbb C}$ or
\item[(iii)] The unit disc $\Delta$.
\end{enumerate}

We remarked earlier that $\mii$ is not compact, so $\mii$ in not conformally
homeomorphic to the Riemann Sphere. It is easy to see that $\mii$ is not
conformally homeomorphic to $\Bbb C$ either. For suppose it were, consider the
function
\be
\phi(\zeta)={1\over f(\zeta)-i}\ ,
\ee
as $\rho>0$ on $\mii$ this is a bounded entire function and hence by Liouville's
theorem $\phi$ (and hence $f$) must be constant. So we are led to
\be
\mii\cong\Delta.
\ee
We may assume from now on that $\zeta$ takes values on the unit disc, $\Delta$.
Next we make use of the asymptotically Melvin nature of the spacetime
 (the following also holds for asymptotically flat solutions):

{\em In coordinates where the point at infinity has a neighbourhood conformally
homeomorphic to the half-disc, with the point at infinity its centre, the
function $f$ should have a simple pole.}

This follows easily by expanding the Melvin solution near infinity.
We therefore map the unit disc to the lower half-plane by means of a M\"obius
transformation, we will be able to extend $f$ to the real axis, where it takes
purely real values. Next we apply Schwarz reflection in this axis to
analytically extend the map to the entire Riemann Sphere. Having defined $f$ 
on the Riemann Sphere allows us to make use of the Valency Theorem.

We consider the pre-image of infinity to work out the valency
 of the map, we have already remarked that $f$ only has a simple pole and 
 so by an application of the Valency Theorem the map $f$ must be 
 univalent, i.e., injective. Hence we
have established that the coordinates $(\rho,z)$ provide a 
diffeomorphism from
$\mii$ to the space $\rho>0$, and therefore $\rho$ and $z$ may be employed as a coordinate
system for the spacetime:
\beq
\b g&=&-V\ddd tt+W(\ddd\phi t+\ddd t\phi)+X\ddd\phi\phi+\Sigma(\ddd\rho\rho+\ddd zz)\nonum\\ &\ &\\
\b A&=&\Phi\dd t+\Psi\dd\phi.
\eq

\section {Determination of the Conformal Factor}
\label{sect:conformal}

We will show in this section that the conformal factor $\Sigma$ decouples from the other equations,
 and for a general harmonic mapping of the type we are discussing
 can be found through quadrature, provided the harmonic mapping equations
 are themselves satisfied. In order to see this we make the dimensional
 reduction from three dimensions to two. We suppose that we have 
already made a dimensional reduction from four dimensions to three
 by exploiting the angular Killing vector, introducing 
Ernst potentials appropriately. The effective Lagrangian then takes the form:
\be
{\cal L}=\sqrt{|\gamma|}\({}^3\!R-2\gamma^{ij}G_{AB}\pp{\phi^A}{x^i}\pp{\phi^B}{x^j}\).
\ee
where the three metric is given by:
\be
\b\gamma=-\rho^2\ddd tt+\Sigma\(\ddd\rho\rho+\ddd zz\).
\ee
Performing the dimensional reduction on the Killing vector
 $\partial/\partial t$, using $\D^2\rho=0$ (w.r.t. the two-dimensional metric) and dropping a
 total divergence we find the effective Lagrangian is
\be
{\cal L}=\sqrt{g^\v_{{}_{\rm II}}}\,g_{{}_{\rm II}}^{\alpha\beta}\(\frc1\Sigma\pp\Sigma{x^\alpha}\pp\rho{x^\beta}
-2\rho\, G_{AB}\pp{\phi^A}{x^\alpha}\pp{\phi^B}{x^\beta}\).
\ee
We have also discarded a term proportional to the Gauss
 curvature of the two dimensional metric. This term makes
 no contribution to the Einstein equations (the two dimensional
 Einstein tensor being trivial) nor does it contribute to the 
harmonic mapping equations. The Einstein equations, derived from variations 
with respect to the metric $\gii$ are easily derived:
\beq
\frc1\Sigma\pp\Sigma z&=&4\rho\, G_{AB}\pp{\phi^A}z\pp{\phi^B}\rho\, ,\\
\frc1\Sigma\pp\Sigma\rho&=&2\rho\, G_{AB}\(\pp{\phi^A}\rho\pp{\phi^B}\rho
-\pp{\phi^A}z\pp{\phi^B}z\).
\eq
Variations with respect to the $\phi^A$ yield the harmonic mapping equation:
\be
\D.\(\rho\,G_{AB}\D\phi^B\)-\frc\rho2\pp{G_{BC}}{\phi^A}\D\phi^B.\D\phi^C=0.
\label{eq:harm}
\ee
This equation can be re-written as
\be
\D.\(\rho\,G_{AB}\D\phi^B\)-\rho\,\Gamma^D_{AC}G_{BD}\D\phi^B.\D\phi^C=0.
\ee
where $\Gamma^D_{AC}$ is the Christoffel symbol derived from the metric $\b G$ on the target space.

Multiplying Eq.~(\ref{eq:harm}) by $\partial\phi^A/\partial z$ leads to
\beq
\ppp z\(\rho\, G_{AB}\)\left[\pp{\phi^A}\rho\pp{\phi^B}\rho-\pp{\phi^A}z\pp{\phi^B}z\right]
&=&2\ppp\rho\(\rho\,G_{AB}\)\pp{\phi^A}z\pp{\phi^B}\rho\nonum\\
&&{}+2\rho\,G_{AB}\left[\pp{^2\phi^B}{\rho^2}+\pp{^2\phi^B}{z^2}\right]
\pp{\phi^A}z.\hspace{1cm}
\eq
It is now a simple matter to calculate the integrability condition for $\Sigma$:
\beq
\h\(\ppp z\ppp\rho \log\Sigma-\ppp\rho\ppp z\log\Sigma\)&=&
\ppp z\(\rho\,G_{AB}\)\left[\pp{\phi^A}\rho\pp{\phi^B}\rho
-\pp{\phi^A}z\pp{\phi^B}z\right]
\nonum\\&&{}
-2\ppp\rho\(\rho\,G_{AB}\)\pp{\phi^A}z\pp{\phi^B}\rho
\nonum\\&&{}
-2\rho\,G_{AB}\left[\pp{^2\phi^B}{\rho^2}+\pp{^2\phi^B}{z^2}\right]
\pp{\phi^A}z
=0.\hspace{1cm}
\eq
We have shown that $\Sigma$ may be found once the harmonic mapping problem is
 solved. Finally we remark that the overall scale of $\Sigma$ is determined by the asymptotic conditions.

\subsection{Boundary Conditions}
\label{sect:bcs1}

In Sect.~\ref{sect:divergence} we proved the divergence result ${}^{(3)}\D^2S\ge0$. The implied metric is given by
\be
\b \gamma=-\rho^2\ddd tt+\Sigma\(\ddd\rho\rho+\ddd zz\).
\ee
In this system nothing depends on $t$. The divergence result can then be written as ${}^{(2)}\D.(\rho\D S)\ge0$. In this section we present appropriate boundary condition that will be
sufficient to make
\be
\int_{\partial\cal M_{\rm II}}\!\!\!\rho\hd\dd S=0.
\ee

As a consequence of Stokes' theorem, this implies ${}^{(3)}\D^2S=0$ throughout the domain of outer communication. From what we learnt in Sect.~\ref{sect:divergence} this proves that $S$ is constant. This constant will be seen to be zero, since the solutions agree at infinity. This then proves uniqueness.

The boundary conditions for the Ernst solution and the $C$-metric are 
presented seperately due to their different behaviour near infinity. Provided that a candidate solution obeys these conditions and the horizon 
structure coincides with that of the $C$-metric, we may deduce that our candidate
solution is described mathematically by either an appropriate Ernst Solution
or $C$-metric.

Having introduced Weyl coordinates to describe the candidate solution we may
evaluate $k$ the modulus of the elliptic functions by Eq.~(\ref{eq:defmod})
from which we can construct $K$ and $K'$ by Eqs.~(\ref{eq:defK}) and 
(\ref{eq:defK'}). We may then use Eq.~(\ref{eq:fwp}) with $2L^2=z_A-z_S$ 
to relate $(\chi,\eta)$ to
the coordinates $(\rho,z)$ with respect to which we may assume the spacetime metric has been expressed. Once we have expressed the solution
with respect to these new coordinates we may use the analysis in 
Sect.~\ref{sect:parameters} to select an appropriate Ernst Solution to act as
the other solution in the uniqueness proof. The vanishing of the boundary
integral will then allow us to conclude that the two solutions are identical.

In these coordinates the integral expression becomes
\be
\int_{\partial\cal M_{\rm II}}\rho\(d\chi\ppp \eta-
d\eta\ppp \chi\)S=0,
\ee
where from Sect.~\ref{sect:divergence}:
\be
S=\frc{{\df X}^2+2(X_1+X_2)({\df E}^2+{\df B}^2)+({\df E}^2+{\df B}^2)^2
+(\df Y+2E_1B_2-2B_1E_2)^2}{4X_1X_2}\ .
\ee

\label{sect:bcs}
\subsubsection{Boundary Conditions for the Ernst Solution Uniqueness Theorem}

We will impose boundary conditions to make the integrand vanish. On the axis $\chi=0$ we impose
\beq
\frc1X\pp X\chi&=&\frc2\chi+O(1);\\
\pp X\eta&=&O\(\chi^{1/2}\);\\
E&=&O\(\chi^{3/2}\);\\
\pp E\chi&=&O\(\chi^{1/2}\);\\
\pp E\eta&=&O\(\chi^{1/2}\);\\
B&=&O\(\chi^{3/2}\);\\
\pp B\chi&=&O\(\chi^{3/2}\);\\
\pp B\eta&=&O\(\chi^{1/2}\);\\
Y&=&O\(\chi^{5/2}\);\\
\pp Y\chi+2\(E\pp B\chi-B\pp E\chi\)&=&O\(\chi^{3/2}\);\\
\pp Y\eta+2\(E\pp B\eta-B\pp E\eta\)&=&O\(\chi^{1/2}\).
\eq
On the other axis we insist
\beq
\frc1X\pp Xu&=&\frc2u+O(1);\\
\pp X\eta&=&O\(u^{1/2}\);\\
E&=&\frc 2{B_0+8D/q}+O\(u^{3/2}\);\\
\pp Eu&=&O\(u^{1/2}\);\\
\pp E\eta&=&O\(u^{1/2}\);\\
B&=&O\(u^{3/2}\);\\
\pp Bu&=&O\(u^{1/2}\);\\
\pp B\eta&=&O\(u^{1/2}\);\\
Y&=&O\(u^{5/2}\);\\
\pp Yu+2\(E\pp Bu-B\pp Eu\)&=&O\(u^{3/2}\);\\
\pp Y\eta+2\(E\pp B\eta -B\pp E\eta\)&=&O\(u^{1/2}\).
\eq
with $u=K-\chi$. The condition on the fields as one approaches infinity is given by
\beq
X&=&\frc{4}{B_0{}^4L^2\sin^2\theta}\frc1R+O\(\frc1{R^2}\);\\
\frc1X\pp XR&=&-\frc1R+O\(\frc1{R^2}\);\\
\pp X\theta&=&O\(\frc1R\);\\
E&=&\frc2{B_0}-\frc2{B_0{}^3L^2\sin^2\theta}\,\frc1R+O\(\frc1{R^2}\);\\
\pp ER&=&\frc2{B_0{}^3L^2\sin^2\theta}\,\frc1{R^2}+O\(\frc1{R^3}\);\\
\pp E\theta&=&O\(\frc1R\);\\
B&=&O\(\frc1{R^2}\);\\
\pp BR&=&O\(\frc1{R^3}\);\\
\pp B\theta&=&O\(\frc1R\);\\
\pp YR+2\(E\pp BR-B\pp ER\)&=&O\(\frc1{R^4}\);\\
\pp Y\theta+2\(E\pp B\theta-B\pp E\theta\)&=&O\(\frc1R\).
\eq
Near infinity we have set $\chi=R^{-1/2}\sin\theta$ and $\eta=R^{-1/2}\cos\theta$.

The boundary conditions on the horizons are particularly simple, we require
the fields $(X,Y,E,B)$ to be regular and that $X>0$ except where the axis and horizon meet. As $X>0$ on this section of the boundary
$S$ will be well-behaved and hence
\be
\int_{0}^{K}\rho\,d\chi{\pp S\eta}=0
\ee
as a result of $\rho=0$.

\subsubsection{Boundary Conditions for the $C$-metric Uniqueness Theorem}
\label{sect:Cboundary}

The boundary conditions will need to impose for the $C$-metric uniqueness result differ from those we required for the Ernst solution (with $B_0=0$) only in the condition at infinity. We require
\beq
X&=&4L^2\sin^2\theta\,R+O(1);\\
\frc1X\pp XR&=&\frc1R+O\(\frc1{R^2}\);\\
\pp X\theta&=&O\(R\);\\
E&=&O\(\frc1R\);\\
\pp ER&=&O\(\frc1{R^2}\);\\
\pp E\theta&=&O\(\frc1R\);\\
B&=&O\(\frc1{R^2}\);\\
\pp BR&=&O\(\frc1{R^3}\);\\
\pp B\theta&=&O\(\frc1R\);\\
\pp YR+2\(E\pp BR-B\pp ER\)&=&O\(\frc1{R^4}\);\\
\pp Y\theta+2\(E\pp B\theta-B\pp E\theta\)&=&O\(\frc1R\).
\eq

These condition are sufficient to cause the appropriate boundary integral to
 vanish and allow us to deduce the uniqueness of the $C$-metric. 
We might remark that the conical singularity that runs along one or both parts
 of the axis in this solution causes us no problem once we pass to the space of
 orbits $\mii$. Strictly we need to require that the range of  the angular
 coordinate of our $C$-metric solution be chosen match that of our candidate
 solution at some point on the axis. The same remark may be made for the Ernst 
solution uniqueness theorem proved in the previous subsection.

\subsection{Summary and Conclusion}
\label{sect:sc}

We have studied the problem of extending the black hole uniqueness proofs to
cover accelerating black holes as represented by the $C$-metric. In addition,
we have considered the case where the acceleration has a physical motivating
force in the form of a cosmological magnetic field; this situation
being modelled by the Ernst Solution. By understanding these solutions to
Einstein-Maxwell theory we have constructed a new set of coordinates that
turn out to be intimately connected with the theory of elliptic functions 
and integrals. At first sight this appears to be a troublesome complication,
however the elliptic functions are naturally defined on a one-parameter set
of rectangles that in some sense are as natural as defining trigonometric 
functions
on a range of $0$ to $2\pi$. The uniqueness proof makes good use of these
standard rectangles and ultimately the divergence integral that
finishes off the proof is over the boundary of one of them.

We also showed how the use of Riemann surface theory assists us to prove the
validity of introducing Weyl coordinates in the Domain of Outer Communication.
We are fortunate in that the Riemann Mapping Theorem for Riemann surfaces
does much of the hard work. We also made good use of the Valency Theorem for
compact Riemann surfaces, this is presented as an alternative to using the Morse theory
that is often employed to prove this step in the uniqueness theorems.

After showing how to determine the conformal factor for the induced 
two-dimensional
metric for any sigma-model, we presented a proof of the positivity of
the divergence required to finish off the uniqueness results. This made use of
the construction of the Bergmann metric from Sect.~\ref{sect:Poincare}. It
contrasts in style with both existing proofs due to Bunting \cite{Bunting}
and Mazur \cite{Mazur,Mazur2}.

We then discussed the boundary conditions required to make the appropriate
boundary integral vanish. Fortunately the boundary conditions are as good
as one could hope. They are able to distinguish between different Ernst
solutions (as indeed they must) and yet they are not overly restrictive. The asymptotically
Melvin nature of these solutions we consider uniquely determines the
cosmological magnetic field parameter, $B_0$ at infinity. The condition on
the axis determine what we have called $D/q$. The boundary conditions
place no other consistency requirements. The known parameters then determine
the other parameters: mass, $m$, charge $g$ and acceleration $A$ for the
solution.

Some authors believe that the semi-classical process of black hole monopole pair creation can be mediated by instantons constructed from the Ernst solution, \cite{Garyfields,Garf}. The uniqueness theorems we have established in the chapter foils
a possible objection to the interpretation of these instantons as mediating
a topology changing semi-classical process. By showing that the instanton is
unique we rule out the possibility that the dominant contribution to the
path integral be given by another exact solution. Had another solution existed
we would have needed to ask which classical action were largest and possibly
which contour we would have to take.

\appendix
\section{Elliptic Integrals and Functions}
\label{sect:elliptic}

In the text we present a black hole uniqueness theorem for 
the Ernst Solution. It turns out that elliptic integrals and the Weierstrass and
 Jacobi elliptic functions provide a valuable tool in establishing that result.
 In this appendix we will establish our
 conventions and collect together most of the general mathematical results
concerning these functions which we will use. These results are discussed in Whittaker and Watson \cite{W&W}, where proofs may be found.

Just as the sine and cosine functions can be regarded as functions on a circle, 
when we have a doubly periodic function we may form the quotient of $\Bbb C$ by 
its period set. Let us call the period set $\Omega$. When we quotient $\Bbb C$ by 
the lattice $\Omega$ we produce a torus. In general two different lattices 
produce conformally inequivalent tori. For our purposes we will only need to consider
 lattices of the form $2\omega_1{\Bbb Z}+2\omega_3{\Bbb Z}$, where $\omega_1$ is
 real and $\omega_3$ is purely imaginary.

The Valency Theorem states that a non-constant analytic function between two compact
 Riemann Surfaces is exactly $n$-1 for some $n$, which we call its {\em valency}.
We shall be applying this result when one of the compact Riemann surfaces is
 the Riemann Sphere and the other one is of these tori.

The first function we will need is the Weierstrass $\wp$-function, this is a doubly
 periodic meromorphic function with a second order pole at the origin, defined by
\be
\wp_\Omega(\zeta)={1\over \zeta^2}+\sum_{\omega\in\Omega\setminus
\{0\}}\({1\over(\zeta-\omega)^2}-{1\over\omega^2}\).
\ee
We will define $\omega_2=\omega_1+\omega_3$ and $e_i=\wp(\omega_i)$. The $\wp$-function
 obeys the differential equation:
\be
\wp'(\zeta)^2=4(\wp(\zeta)-e_1)(\wp(\zeta)-e_2)(\wp(\zeta)-e_3).
\ee
This is easily established by noting that the ratio of the LHS and the RHS has 
no poles
 and is therefore, by the Valency Theorem, constant, the constant is determined by
 examining what happens as $\zeta\rightarrow0$. Looking at this limit we see
\be
e_1+e_2+e_3=0.
\ee
The differential equation is then given by
\be
\wp'(\zeta)^2=4\wp(\zeta)^3-g_2\wp(\zeta)-g_3.
\ee
We will call $g_2$ and $g_3$ the {\em invariants} of the $\wp$-function. Note that
\beq
e_1e_2+e_2e_3+e_3e_1&=&-\sq g_2\qquad\mbox{and}\\
e_1e_2e_3&=&\sq g_3.
\eq
Therefore
\be
(e_1-e_2)^2+(e_2-e_3)^2+(e_3-e_1)^2=\mbox{$\frac32$}g_2
\label{eq:esquared}
\ee
and
\be
(e_1-e_2)^2(e_2-e_3)^2(e_3-e_1)^2=\frc{g_2{}^3-27g_3{}^2}{16}.
\label{eq:ecubed}
\ee
Integrating the differential equation we find the elliptic integral
\be
\zeta=\int^\infty_z\frc{dt}{\sqrt{4t^3-g_2t-g_3}}
\ee
has the solution $z=\wp(\zeta)$. We also point out that the Weierstrass $\wp$-function has the scaling property:
\be
\wp(M\zeta;M\Omega)=M^{-2}\wp(\zeta;\Omega).
\ee
The invariants of $\wp$ for $M\Omega$ being $g_2'=M^{-4}g_2$, $g_3'=M^{-6}g_3$.

We may evaluate elliptic integrals of the form
\be
\int_{x_0}^x\frc{dt}{\sqrt{f(t)}}
\ee
for any quartic $f(t)=a_0t^4+4a_1t^3+6a_2t^2+4a_3t+a_4$ with a root $x_0$, in terms of an
 appropriate $\wp$ function. The invariants of which are given by
\beq
g_2&=&a_0a_4-4a_1a_3+3a_2{}^2
\label{eq:invt1}
\\
g_3&=&a_0a_2a_4+2a_1a_2a_3-a_2{}^3-a_0a_3{}^2-a_1{}^2a_4.
\label{eq:invt2}
\eq
The result being:
\be
x-x_0=\frc{f'(x_0)}{4\(\wp(\zeta)-\frac1{24}f''(x_0)\)}
\ee
and
\be
\sqrt{f(x)}=\frc{-f'(x_0)\wp'(\zeta)}{4\(\wp(\zeta)-\frac1{24}f''(x_0)\)^2}.
\ee

The Weierstrass function will be extremely useful in what follows. It is also highly useful to 
define the Jacobi elliptic functions that in some sense generalize the trigonometric functions.
\beq
\sn\zeta&=&\sqrt{\frc{e_1-e_3}{\wp(\zeta/M)-e_3}},\\
\cn\zeta&=&\sqrt{\frc{\wp(\zeta/M)-e_1}{\wp(\zeta/M)-e_3}},\\
\dn\zeta&=&\sqrt{\frc{\wp(\zeta/M)-e_2}{\wp(\zeta/M)-e_3}},\\
\eq
with $M=\sqrt{e_1-e_3}$.  These functions obey certain algebraic and differential identities. We 
introduce the {\em modulus\/}, $k$ defined by
\be
k^2=\frc{e_2-e_3}{e_1-e_3}
\ee
and the {\em complementary modulus\/}, $k'$ defined by $k'^2=1-k^2$. For the situation we will be considering the modulus will be real-valued and in the range $[0,1]$. The algebraic identities we need are
\beq
\cn^2\zeta&=&1-\sn^2\zeta,\\
\dn^2\zeta&=&1-k^2\sn^2\zeta.
\eq
Whilst the derivatives are given by
\beq
\frc d{d\zeta}\sn\zeta&=&\cn\zeta\dn\zeta,\\
\frc d{d\zeta}\cn\zeta&=&-\sn\zeta\dn\zeta,\\
\frc d{d\zeta}\dn\zeta&=&-k^2\sn\zeta\cn\zeta.
\eq
The functions $\sn\zeta$ and $\cn\zeta$ are periodic with periods $4K$ whilst that of $\dn\zeta$ has period $2K$ where
\be
K=\int_0^1\frc{dt}{\sqrt{(1-t^2)(1-k^2t^2)}}.
\label{eq:defK}
\ee
It is useful to define the analogous quantity associated with the complementary modulus, namely:
\be
K'=\int_0^1\frc{dt}{\sqrt{(1-t^2)(1-k'^2t^2)}}.
\label{eq:defK'}
\ee
After rescaling so that $\epsilon_i=e_i/M^2$ we find that
\beq
\epsilon_1&=&\frac13(1+k'^2),\\
\epsilon_2&=&\frac13(k^2-k'^2),\\
\epsilon_3&=&-\frac13(1+k^2).
\eq
The Jacobi functions obey the addition rules:
\beq
\sn(\zeta+C)&=&\frc{\sn\zeta\cn C\dn C+\sn C\cn\zeta\dn\zeta}{1-k^2\sn^2\zeta\sn^2 C}\ ,\\
\cn(\zeta+C)&=&\frc{\cn\zeta\cn C-\sn\zeta\sn C\dn\zeta\dn C}{1-k^2\sn^2\zeta\sn^2 C}\ ,\\
\dn(\zeta+C)&=&\frc{\dn\zeta\dn C-k^2\sn\zeta\sn C\cn\zeta\cn C}{1-k^2\sn^2\zeta\sn^2 C}\ .
\eq
In particular when $C=K$ we can use $\sn K=1$, $\cn K=0$, $\dn K=k'$ together with $\sn 0=0$, $\cn 0=\dn 0=1$ and 
the fact that $\sn\zeta$ is an odd function of $\zeta$ whilst both $\cn\zeta$ and $\dn\zeta$ are 
even to deduce
\beq
\sn(K-\zeta)&=&\frc{\cn\zeta}{\dn\zeta}\ ,\\
\cn(K-\zeta)&=&\frc{k'\sn\zeta}{\dn\zeta}\ ,\\
\dn(K-\zeta)&=&\frc{k'}{\dn\zeta}\ .
\eq
Although these formulae are valid for all complex values of $\zeta$ it will be convenient to 
write $\zeta=\chi+i\eta$ and to be able to decompose the elliptic functions into real and 
imaginary parts in terms of functions of $\chi$ and $\eta$. The addition formulae allow
 us to do this provided we know the values of the functions evaluated on a purely
 imaginary argument. For this we use Jacobi's imaginary transform:
\beq
\sn i\eta&=&i\,\frc{\sn\eta}{\cn\eta},\\
\cn i\eta&=&\frc1{\cn\eta},\\
\dn i\eta&=&\frc{\dn\eta}{\cn\eta},
\eq
where importantly the elliptic functions on the RHS of each of the above equations
 is with modulus $k'$. For brevity we will always regard the elliptic functions as
 being with modulus $k$ unless the argument is $\eta$ when it should be understood
 that the modulus is $k'$. This should not cause confusion in this paper as we
 will be doing few manipulations involving Jacobi elliptic functions with respect to 
the complementary modulus.

We will need to expand $\sn\zeta$, $\cn\zeta$ and $\dn\zeta$ for small values of the 
argument. We find
\beq
\sn\zeta&=&\zeta-\frac16(1+k^2)\zeta^3+O\(\zeta^5\),\\
\cn\zeta&=&1-\frac12\zeta^2+O\(\zeta^4\),\\
\dn\zeta&=&1-\frac12k^2\zeta^2+O\(\zeta^4\).
\eq

Finally we note that the $\wp$-function with $\omega_1=K$ and $\omega_3=iK'$ may be expressed in terms
 of the Jacobi functions by
\be
\wp(\zeta)=-\frc{1+k^2}3+\frc1{\sn^2\zeta}=\frc{\cn^2\zeta+\frac13(1+k'^2)\sn^2\zeta}{\sn^2\zeta}.
\ee


\end{document}